\documentclass[twocolumn]{aastex63}

\usepackage{amsmath}	
\usepackage{hyperref}
\usepackage{xcolor}

\newcommand{\Mdot}{\dot{M}}
\newcommand{\Edot}{\dot{E}}
\newcommand{\Ledd}{L_\text{Edd}}
\newcommand{\Lcr}{L_\text{cr}}
\newcommand{\Linf}{L^\infty}
\newcommand{\mach}{\mathcal{M}}
\newcommand{\ph}{_\text{ph}}
\newcommand{\co}{_\text{co}}
\newcommand{\fx}{_\text{fx}}
\newcommand{\Teff}{T_\text{eff}}

\shorttitle{Expanded structures in PRE bursts}
\shortauthors{Guichandut et al.}

\graphicspath{{./}{figures/}}

\begin{document}

\title{Expanded atmospheres and winds in Type I X-ray bursts from accreting neutron stars}

\correspondingauthor{Simon Guichandut}
\email{simon.guichandut@mcgill.ca}

\author{Simon Guichandut}
\affiliation{Department of Physics and McGill Space Institute, McGill University, 3600 rue University, Montreal, QC, H3A 2T8, Canada}

\author{Andrew Cumming}
\affiliation{Department of Physics and McGill Space Institute, McGill University, 3600 rue University, Montreal, QC, H3A 2T8, Canada}

\author{Maurizio Falanga}
\affiliation{International Space Science Institute (ISSI), Hallerstrasse 6, 3012 Bern, Switzerland}

\author{Zhaosheng Li}   
\affiliation{Key Laboratory of Stars and Interstellar Medium, Xiangtan University, Xiangtan 411105, Hunan, P.R. China}

\author{Michael Zamfir}
\affiliation{Department of Physics and McGill Space Institute, McGill University, 3600 rue University, Montreal, QC, H3A 2T8, Canada}

\begin{abstract}
We calculate steady-state models of radiation-driven super-Eddington winds and static expanded envelopes of neutron stars caused by high luminosities in type I X-ray bursts. We use flux-limited diffusion to model the transition from optically thick to optically thin, and include effects of general relativity, allowing us to study the photospheric radius close to the star as the hydrostatic atmosphere evolves into a wind. We find that the photospheric radius evolves monotonically from static envelopes ($r_{\rm ph}\lesssim 50$--$70\ {\rm km}$) to winds ($r_{\rm ph}\approx 100$--$1000\ {\rm km}$). Photospheric radii of less than 100 km, as observed in most photospheric radius expansion bursts, can be explained by static envelopes, but only in a narrow range of luminosity. In most bursts, we would expect the luminosity to increase further, leading to a wind with photospheric radius ${\gtrsim}\, 100\ {\rm km}$. 
In the contraction phase, the expanded envelope solutions show that the photosphere is still ${\approx}\,1\ {\rm km}$ above the surface when the effective temperature is only 3\% away from its maximum value. This is a possible systematic uncertainty when interpreting the measured Eddington fluxes from bursts at touchdown. We also discuss the applicability of steady-state models to describe the dynamics of bursts. In particular, we show that the sub to super-Eddington transition during the burst rise is rapid enough that static models are not appropriate. Finally, we analyze the strength of spectral shifts in our models. Expected shifts at the photosphere are dominated by gravitational redshift, and are therefore predicted to be less than a few percent. 
\end{abstract}


\section{Introduction}\label{sec:intro}

Type I X-ray bursts are transient astronomical events that result from thermonuclear runaways on the surface of accreting neutron stars (see \citealt{Galloway2017} for a comprehensive review). By making the neutron star shine brightly in X-rays for many seconds to minutes, these bursts provide a unique opportunity to study the properties of the star directly. A subset of bursts produce high enough fluxes that hydrostatic equilibrium on the surface is lifted, leading to an expansion of the star's photosphere \citep{Lewin1993}. In these \textit{photospheric radius expansion} (PRE) bursts, the observed blackbody temperature quickly drops from its initial state of ${\sim}\,2$--$3$ keV to ${\lesssim}\,0.5$--1 keV, then slowly rises back to its initial state during the burst decay, as the photosphere falls back down to the surface. The inferred blackbody or photospheric radius correspondingly rises and falls back down again. Traditionally, observations of PRE bursts show an artificial dip in flux as the peak energy of the emission moves out of the spectral band of the detector, although recently \textit{NICER} (\textit{Neutron Star Interior Composition Explorer}), a modern X-ray telescope with a soft X-ray response, has been able to track the complete spectral evolution of PRE bursts \citep{Keek2018a}. In this paper, we revisit models of the neutron star envelope as it expands and evolves into a wind.

\subsection{Observational Motivation}\label{subsec:observational_motivation}

A number of recent observations motivate our work. The first is the large sample of PRE bursts now available, including extreme events with very dramatic radius expansion, as well as long-lasting super-Eddington phases. 
Of the more than 7000 bursts from 85 sources in the MINBAR (\textit{Multi-Instrument Burst Achive}) catalog \citep{Galloway2020}, about one fifth exhibit PRE. Most show moderate expansion in which the photosphere expands to tens of kilometres above the stellar surface, but in rare \textit{superexpansion} bursts, the photosphere reaches a thousand kilometres in radius or more \citep{IntZand2010}. Superexpansion bursts occur in ultracompact binaries in which the neutron star accretes helium at low rates, giving rise to long duration bursts with Eddington phases lasting for minutes \citep{intZand2005}. An extreme example is the 4U 1820-30 superburst, which was at the Eddington limit for approximately 20 minutes \citep{Strohmayer2002,IntZand2010}. \cite{IntZand2010} noted that whereas the overall duration of the Eddington phase scales with the burst fluence, so that more energetic bursts have longer Eddington phases, the duration of the superexpansion phase is always short, approximately a few seconds. The reason for this is not clear, but could be the result of ejection of a shell of material during the onset of a wind \citep{IntZand2010}, or the timescale for the wind to be polluted with heavy nuclear burning ashes \citep{YuWeinberg2018}.

The second motivation is to address systematic effects in using PRE bursts to determine neutron star radii, and thereby constrain the dense matter equation of state (see \citealt{OzelARAA2016} for a review). The expansion of the photosphere at the Eddington luminosity provides a means to place joint constraints on both mass and radius of the star \citep{Lewin1993,Ozel2016}. The principal way of doing this is by interpreting the peak of the blackbody temperature curve, after the expansion phase, as the \textit{touchdown} point, i.e. the moment where the atmosphere collapses back down to the surface and the photospheric radius is equal to the neutron star radius. However, the systematics of the photosphere's evolution and touchdown point are not well understood, as discussed for example by \cite{Galloway2008touchdown} and \citet{Steiner2010}. \cite{Steiner2010} found that allowing the photospheric radius to be larger than the neutron star radius at touchdown gave a much larger number of physical solutions for the mass and radius. \cite{Ozel2016} however argued against this interpretation and showed that including rotational corrections and the temperature-dependence of the opacity alleviates this issue (see \citealt{Suleimanov2020} for further discussion). Nonetheless, the extent to which the photosphere remains elevated at the touchdown point is still an open question.

Third, recent observational evidence of spectral edges and lines in type I X-ray burst spectra \citep{IntZand2010,Kajava2017,Li2018,Strohmayer2019} suggests that heavy elements are present near the photosphere during the expansion phase. This is of interest since type I X-ray bursts are sites of heavy element production by the rp-process (e.g.~\citealt{Schatz2003}), and measuring the gravitational redshift of spectral features would provide an important constraint on the stellar mass and emission radius, and therefore the neutron star mass-radius relation. \cite{Weinberg2006} and \cite{YuWeinberg2018} show that convection at the onset of the burst brings nuclear burning ashes to low enough column depths that they could be ejected by a wind. Both absorption/emission lines and photoionization edges are expected to be present in the burst spectra, though their observational signature will be subject to complicated effects such as rotational broadening in the case of lines \citep{Chang2005} and the exact composition of the outflow in the case of edges \citep{Weinberg2006}. 

\citet{Strohmayer2019} presented observations of bursts from 4U~1820-30 with \textit{NICER} that showed emission and absorption lines. They found three lines that were blueshifted by the same factor of ${\approx}\,1.046$ in a set of bursts with photospheric radii of ${\approx}\,100\ {\rm km}$ compared to weaker bursts with photospheric radii of ${\approx}\,75\ {\rm km}$. This is consistent with line emission associated with heavy elements in the wind in the sense that the lines in the weaker bursts should be produced closer to the star, and thus have stronger gravitational redshift, and in addition have smaller wind velocities and therefore weaker Doppler blueshifts. \cite{Strohmayer2019} suggested that both of these effects could work together to create the observed shifts, although they noted that the observed shifts are perhaps larger than expected from wind models. The gravitational redshift difference corresponding to a change in emission radius from 75 to $100\ {\rm km}$ is only $\approx 1$\%, with a $\lesssim 1$\% addition effect from Doppler shift due to the wind velocity.

\subsection{Static Envelopes and Winds}

There are two types of solution for the neutron star envelope in which the photosphere moves to large radius,
depending on whether the luminosity at the base of the envelope as seen at infinity, $L_b^\infty$, is larger or smaller than the Eddington luminosity, 
\begin{eqnarray}\label{eq:LEdd}
    \Ledd &\equiv& \frac{4\pi GMc}{\kappa_0}\\&=&3.5\times 10^{38}\ {\rm erg\ s^{-1}}\,\left(\frac{M}{1.4M_\odot}\right)\left(\frac{1}{1+X}\right),\nonumber
\end{eqnarray}
where $\kappa_0 = 0.2\,(1+X)\ {\rm cm^2\ g^{-1}}$ is the constant electron scattering opacity, and $X$ is the hydrogen fraction \citep{Clayton1983}. When $L^\infty_b>L_{\rm Edd}$, the luminosity in excess of Eddington is used to drive mass loss, and a super-Eddington wind forms with radiative luminosity ${\approx}\,\Ledd$, and a mass-loss rate
\begin{equation}
    \Mdot\approx \frac{L_b^\infty-\Ledd}{GM/R}\sim 10^{18}\ {\rm g\ s^{-1}} \left(\frac{L_b^\infty-\Ledd}{\Ledd}\right) \label{eq:mdotprescription}
\end{equation}
\citep{Paczynski1986b}. At luminosities below, but close to, the Eddington luminosity ($0.7\lesssim L_b^\infty/ \Ledd \lesssim 1$), \cite{Paczynski1986a} showed that in general relativity there is a sequence of expanded hydrostatic envelopes which can extend outwards as much as ${\sim}\, 200\ {\rm km}$.

Both of these solutions depend crucially on the fact that at the high temperatures ${\gtrsim}\,10^9\ {\rm K}$ reached in bursts, Klein-Nishina corrections reduce the electron scattering opacity. This leads to a significant increase in the local Eddington luminosity or {\em critical luminosity} 
\begin{equation}\label{eq:Lcrit}
    \Lcr=\frac{4\pi GMc}{\kappa(\rho,T)}\left(1-\frac{2GM}{rc^2}\right)^{-1/2}.
\end{equation}
This means that even when the luminosity is super-Eddington,  it can be well below the critical luminosity at the base of the envelope, allowing the hydrostatic envelope to carry the super-Eddington flux \citep{Hanawa1982}. At larger radii, the temperature and $\Lcr$ drop, so that $L$ approaches and can eventually exceed $\Lcr$, leading to outwards expansion. This leads to a compact geometrically-thin inner envelope in hydrostatic equilibrium that transitions into an extended outer region. In the case of static envelopes, the structure adjusts such that the luminosity at each radius is slightly below the local critical luminosity (to one part in $10^4$; \citealt{Paczynski1986a}), maintaining hydrostatic balance but with a very extended structure. In the case of winds, the luminosity is similarly close to the critical luminosity until the fluid reaches the sound speed, at which point the luminosity becomes super-critical (typically by ${\sim}\,1$\%, e.g.~\citealt{Paczynski1986b}), furthering the acceleration of the material to velocities ${\sim}\,0.01c$.

The burst wind regime has been studied extensively, with different approximations. The first studies calculated steady-state wind solutions assuming Newtonian gravity and optically-thick radiative transfer \citep{Ebisuzaki1983,Kato1983a}. \cite{Quinn1985} improved the treatment of the outer boundary with an approximate form for the transition from optically-thick to optically-thin parts of the wind. More recently, \cite{Herrera2020} carried out a more detailed survey of the available parameter space for these kinds of models, with an emphasis on predicting correlations between photospheric quantities. These studies established the basic features of super-Eddington winds from type I X-ray bursts, namely the radiative luminosity is within a few percent of the Eddington luminosity, the outflow velocity is ${\sim}\,0.01 c$, and photospheric radii ranging from tens of km to ${\gtrsim}\,1000\ {\rm km}$ for the highest mass-loss rates. These conclusions carried over into the works of \cite{Joss1987}, \cite{Titarchuk1994} and \cite{Shaposhnikov2002}, that included a more detailed discussion of the radiative transfer in the wind, and \citet{YuWeinberg2018} who calculated the first time-dependent models.

In contrast to Newtonian models, \cite{Paczynski1986b} showed that in wind models that include general relativity, the photospheric radius is always more than an order of magnitude larger that the neutron star radius, even at low mass-loss rates. This is very different from Newtonian models, in which the photospheric radius reduces smoothly to the neutron star radius at low mass-loss rates. This behavior is consistent with the expanded hydrostatic envelopes that \cite{Paczynski1986a} found when general relativity was included. The results of \cite{Nobili1994}, which included general relativity as well as a more sophisticated treatment of radiative transfer, show a similar result that the photospheric radius is ${\gtrsim}\, 100 \ {\rm km}$ at low mass-loss rates. 

These results suggest that it is crucial to include general relativity if we are interested in photospheric radii ${\lesssim}\,100\ {\rm km}$, as observed in the majority of PRE bursts, or if we are interested in understanding how the photospheric radius behaves as the burst luminosity drops below Eddington and touches down. For example, based on their results, \cite{Paczynski1986b} concluded that a super-Eddington wind should have a photospheric temperature too low to be detected by available X-ray instruments, implying that observed PRE bursts had hydrostatic expanded atmospheres rather than winds. 

In this paper, we study the evolution of the envelope around the transition from hydrostatic to wind, when the photospheric radii are expected to be close to the star.
We include general relativity and use flux-limited diffusion to model the transition from optically thick to optically thin regions, allowing us to extend both static atmosphere and wind solutions out to low optical depths and thereby use a consistent definition of the photosphere in both. This improves on the calculations of \cite{Paczynski1986a} and \cite{Paczynski1986b}, which assumed optically thick radiation transport and had different prescriptions for the photosphere. For static envelopes, \cite{Paczynski1986a} set $\tau=\int \rho \kappa dr=2/3$ at the location where $T = T_{\rm eff}\equiv (L/4\pi r^2\sigma)^{1/4}$, whereas for winds \cite{Paczynski1986b} instead set the optical depth parameter $\tau^\star = \rho\kappa r=3$.

We focus here on light element envelopes. We show results for pure helium envelopes, but also check that pure hydrogen or solar composition models are not substantially different. It is important to point out that heavy elements may play an important role in the radiative transfer in the wind, and may explain the smaller observed photospheric radii compared to light element winds. \cite{IntZand2010} point out the possible importance of line driving from hydrogenic ions of Ni or other heavy elements.
\citet{YuWeinberg2018} performed the first time-dependent calculations of optically-thick Newtonian winds, with a focus on tracking the composition of different elements over time and space, and proposed that heavy element pollution terminates the superexpansion phase \citep{IntZand2010}. While there can be no doubt that heavy elements must be integrated into models of PRE bursts to take advantage of the observations now available, we start in this paper with a self-consistent set of steady-state, light element models that we can use as a basis for future work.

\subsection{Outline of the Paper}

We start in \S\ref{sec:model} by describing the general equations for steady-state flow in general relativity, and how we apply flux-limited diffusion to model the radiative transfer. We then describe the methods we use to calculate wind and static envelope solutions, including the change of variables proposed by \cite{Joss1987} to integrate near the critical point of the wind. In \S\ref{sec:results}, we present the solution profiles, and discuss the location of the photosphere as the base luminosity varies and the maximum and minimum mass-loss rates. In \S\ref{sec:transition}, we discuss the transition between expanded envelopes and winds as the base luminosity rises during a burst, and the applicability of steady-state solutions.
In \S\ref{sec:implications}, we discuss some observational implications of our results, including the expected photospheric radii, spectral shifts, and the behavior of the photosphere near touchdown. We conclude in \S\ref{sec:discussion}.

\section{Model description and methods}\label{sec:model}
In this section, we explain how we obtained our  wind and envelope solutions. In \S\ref{subsec:equations}, we derive the equations for the steady-state flow of an ideal gas, considering general relativity (GR) under a Schwarzschild metric, and radiation transport described by flux-limited diffusion (FLD). In \S\ref{subsec:wind}, we explain our numerical method for obtaining wind solutions that satisfy boundary conditions at the stellar surface and at large radii, and similarly in \S\ref{subsec:envelopes} for envelopes.

\subsection{Steady-state radiation hydrodynamics with FLD}\label{subsec:equations}
For both the wind and static envelope case, we consider a fluid and radiation field in a spherically symmetric Schwarzschild spacetime, which is characterized by the curvature parameter $\zeta=(1-2GM/c^2r)^{1/2}$, where $G$ is the gravitational constant, $c$ is the speed of light, $M$ is the mass of the neutron star and $r$ is the radial coordinate. The parameter
\begin{equation}
    \Psi\equiv\sqrt{\frac{1-2GM/c^2r}{1-v^2/c^2}}=\zeta\gamma \,,
\end{equation}
where $v$ is the velocity and $\gamma$ is the usual Lorentz factor, is often referred to as the \textit{energy parameter} for the flow, written as ``$Y$'' in \citet{Paczynski1986b} and \cite{Thorne1981}, or ``$y$'' in \cite{Park2006}.

In steady-state, the relativistic radiation hydrodynamics equations can be manipulated to yield conservation equations for mass and energy,
\begin{align}
    &\dot{M}=4\pi r^2\rho v\Psi \label{eq:Mdot}\,,\\
    &\dot{E}=L\Psi^2\left(1+\frac{v^2}{c^2}\right)+\dot{M}\Psi\left(c^2+\frac{P+U}{\rho}\right) \label{eq:Edot}\,,
\end{align}
and a momentum equation for the fluid and radiation\,,
\begin{equation}
    (\rho c^2+P_\text{g}+U_\text{g})\frac{d\ln\Psi}{dr}+\frac{dP_\text{g}}{dr}-\frac{1}{c\Psi}\rho\kappa F=0\label{eq:momentum}\,,
\end{equation}
where $\dot{M}$ and $\dot{E}$ are the mass and energy-loss rates, $\rho$ is the rest-mass density, $F$ is the local (or comoving) radiative flux and $L=4\pi r^2F$ is the local luminosity. The total pressure $P=P_g+P_R$ and energy $U=U_g+U_R$ are the sum of the gas and radiation contributions. In Appendix \ref{sec:derivations}, we show how to obtain these equations from the time-dependent hydrodynamics equations.

For the electron scattering opacity, we use the fitting formula\footnote{\cite{Poutanen2017} presents a more accurate version of this formula, but also shows that Eq.~(\ref{eq:kappa}) is accurate to a few percent (see Fig.~2 of \citealt{Poutanen2017}) which is adequate for our purposes.} from \cite{Paczynski1983},
\begin{equation}
    \label{eq:kappa}
    \kappa=\kappa_0\left[1+\left(\frac{T}{4.5\times 10^8\,\text{K}}\right)^{0.86}\right]^{-1} \,,
\end{equation}
where $\kappa_0=0.2(1+X)$ is the classical scattering opacity from the Thomson cross-section \citep{Clayton1983}. When the local temperature of the gas $T$, becomes large, the cross-section is reduced by Klein-Nishina corrections. \citet{Paczynski1983} also provides a density correction to $\kappa_0$ from electron degeneracy, which we can safely ignore at the densities in our solutions. We have also verified a posteriori that free-free opacity was not important anywhere in our models. Even near the base where the densities are large ($\rho\sim10^4$ g cm$^{-3}$), the temperatures are large enough ($T\sim10^9$ K) that electron-scattering dominates \citep{Schatz1999}.

We assume an ideal monatomic gas equation of state, with pressure and internal energy
\begin{equation}
    P_g = \frac{kT\rho}{\mu m_p} \qquad;\qquad U_g=\frac{3}{2}P_g \label{eq:eos}\,,
\end{equation}
where $k$ is the Boltzmann constant, $m_p$ is the proton mass and $\mu$ is the mean molecular weight. Note that we treat $\mu$ as a constant of the model, meaning the composition of the gas is fixed. We write the radiation  energy density as $U_R=aT^4$ where $a$ is the radiation constant, and we will define the radiation pressure $P_R$ later in this section. We use the usual ratio $\beta=P_g/P$ to track the relative importance of these pressures throughout the flow. As discussed by \cite{Quinn1985} and \cite{Joss1987}, at moderate to high optical depths, LTE applies and the gas and radiation can be described with a single temperature $T$ (even though the opacity is scattering dominated, Compton scattering is able to keep the photons and gas at the same temperature; \citealt{Joss1987}). In regions of low optical depth, $T$ measures the radiation energy density via $T^4=U_R/a$. We still use eqs.~\eqref{eq:kappa} and \eqref{eq:eos} in these regions since the gas pressure is negligible ($\beta\ll 1$) and $\kappa\approx\kappa_0$, independent of temperature.

We model the transition between optically thick and optically thin regions using FLD as described by \citet{Levermore1981}, but with added factors of $\Psi$ to account for general relativity and produce the correct limits (see \citealt{Rahman2019} for a similar approach to neutrino transport in GR). The radiative flux is given by
\begin{equation}\label{eq:FLD_flux}
    F=\frac{-\lambda c\nabla(\Psi^4U_R)}{\kappa\rho\Psi^3}\,,
\end{equation}
where the interpolating factor $\lambda$ is
\begin{equation}\label{eq:lambda_R}
    \lambda=\frac{2+R}{6+3R+R^2} \quad;\quad R=\frac{1}{\kappa\rho\Psi^3}\frac{\vert\nabla(\Psi^4U_R)\vert}{U_R}\,.
\end{equation}
In the optically thick regions, a short mean free path results in $R\rightarrow 0$, $\lambda\rightarrow 1/3$, and Eq.~\eqref{eq:FLD_flux} becomes the standard photon diffusion equation, with additional factors of $\Psi$ because of GR (see \citealt{Paczynski1986b} and \citealt{Flammang1984}). In the optically thin regions, $R$ becomes large such that $\lambda\rightarrow 1/R$ and $F\rightarrow cU_R$, the correct photon streaming limit. Note that this limit gives an analytical expression for the radiation temperature in the optically thin limit,
\begin{equation}\label{eq:T_thin}
    T_{\lambda\rightarrow 0}=\left(\frac{L}{4\pi r^2 ac}\right)^{1/4},
\end{equation}
which is useful since the luminosity is nearly constant there. In the transition between the two regions, it is not possible to exactly describe the radiation without explicitly solving the full radiative transfer equations, but the smooth and monotonic transition controlled by $\lambda$ should be satisfactory. At any point, if the flux and temperature are known, $\lambda$ can be calculated by solving
\begin{equation}
    6\lambda^2+\lambda(3x-2)+x(x-1)=0\quad;\quad x\equiv\frac{F}{cU_R}.\label{eq:FLD_lambda_x}
\end{equation}
FLD can also be used to interpolate the radiation pressure tensor with the flux limiter $\lambda$ \citep{Levermore1984}. In 1D, this gives a simple relation for the radiation pressure in the radial direction,
\begin{equation}
    P_R=(\lambda+\lambda^2R^2)U_R\,.\label{eq:radiation_pressure}
\end{equation}
This has limits of $P_R=U_R/3$ in optically thick regions, where the radiation is isotropic, and $P_R=U_R$ in optically thin regions, where the radiation is beamed in the radial direction\footnote{Previous work such as \citet{Quinn1985} took the radiation pressure as the optically thick expression even in optically thin regions, and explained that this only resulted in errors of order $v/c$. We made models with both prescriptions for $P_R$, and while it is true that they give similar qualitative results, the accumulation of errors displaces the photosphere by up to 5\%.}.

By combining the fluid equations (eq.~[\ref{eq:Mdot}]--[\ref{eq:momentum}]) with the radiative flux limited diffusion prescription (eq.~[\ref{eq:FLD_flux}]) and the equation of state (eq.~[\ref{eq:eos}]), the equations of structure can be derived in the form of three coupled first order ordinary differential equations for the velocity, density and temperature of the gas. These are
\begin{gather}
    \frac{d\ln v}{d\ln r}(c_s^2-v^2A)\gamma^2=\frac{GM}{r\zeta^2}\left(1+\frac{1}{2}\frac{c_s^2}{c^2}\right)-2c_s^2-C\label{eq:dvdr}\,,\\
    \frac{d\ln\rho}{d\ln r}(c_s^2-v^2A)=\left(2v^2-\frac{GM}{r\Psi^2}\right)A+C \label{eq:drhodr}\,,\\
    \frac{d\ln T}{d\ln r}=-T^*-\frac{GM}{c^2\zeta^2r}-\frac{\gamma^2v^2}{c^2}\frac{d\ln v}{d\ln r}\,, \label{eq:dTdr}
\end{gather}
with the sound speed defined by $c_s^2\equiv P_g/\rho$, and the parameters
\begin{align}
    \label{eq:constants}
    &T^*=\frac{\kappa\rho rF}{4acT^4\lambda\Psi}=\frac{1}{\lambda\Psi}\frac{L}{L_\text{Edd}}\frac{\kappa}{\kappa_0}\frac{GM}{4r}\frac{\rho}{U_R}\,,\nonumber\\
    &A=1+\frac{3}{2}\frac{c_s^2}{c^2}\nonumber\,,\\
    &C=\frac{1}{\Psi}\frac{L}{L_\text{Edd}}\frac{\kappa}{\kappa_0}\frac{GM}{r}\left(1+\frac{\beta}{12\lambda(1-\beta)}\right)\,.
\end{align}
Eq.~\eqref{eq:dvdr}-\eqref{eq:constants} allow us to extend the calculations of \citet{Paczynski1986b} to optically thin regions. In the optically thick limit ($\lambda=1/3$), they reduce to the exact equations written in this previous paper.

\subsection{Winds}\label{subsec:wind}

For winds, we look for solutions to equations \eqref{eq:dvdr} and \eqref{eq:dTdr} that have a small velocity near the surface of the neutron star and that continuously accelerate to large radii. These solutions have an important location called the \textit{sonic point}, where the fluid goes from being subsonic to supersonic. This is due to the sound speed $c_s$ decreasing with temperature. This point always appears as a singularity in the velocity gradient equation, no matter the equation of state \citep{LamersCassinelli}. In our Eq.~\eqref{eq:dvdr} (also eq.~[\ref{eq:drhodr}]), the singularity occurs when $v=v_s\equiv c_s/\sqrt{A}$. Note that this is not exactly the sound speed due to small GR corrections of order $(c_s/c)^2$ (see \citealt{Paczynski1986b}). In order for the solutions to smoothly pass through the sonic point, the right-hand side of Eq.~\eqref{eq:dvdr} also needs to vanish, which defines the sonic point radius $r_s$, and its temperature $T_s$.

To avoid numerical difficulties around $r_s$, we adapt the approach of \cite{Joss1987} (who solved the Newtonian equations), and introduce a new dimensionless variable
\begin{equation}
    \label{eq:phi}
    \Phi=A^{1/2}\mach+A^{-1/2}\mach^{-1} \,,
\end{equation}
where $\mach=v/c_s$ is the Mach number. $\Phi$ has a value of exactly 2 at the sonic point.  The gradient is given by
\begin{align}
    \frac{d\Phi}{dr}=&\frac{(A\mach^2-1)(3c_s^2-2Ac^2)}{4\mach A^{3/2}c^2r}\frac{d\ln T}{d\ln r}\nonumber\\
    -&\frac{c_s^2-Av^2}{vc_sr\sqrt{A}}\frac{d\ln v}{d\ln r} \,.
    \label{eq:dphidr}
\end{align}
When substituted into equation \eqref{eq:dvdr}, the singularity at the sonic point cancels, allowing a smooth integration through $r_s$.  As for equation \eqref{eq:dTdr}, the last term may easily be ignored since $v^2\ll c^2$.

We constructed wind solutions for every value of the mass-loss rate $\Mdot$ with the following method. We first set a trial value for the energy-loss rate $\Edot$, which allows us to solve for the luminosity at any point $r$ given values for $T$ and $v$ using Eq.~\eqref{eq:Mdot}-\eqref{eq:Edot}. We then choose a trial value for the sonic point location $r_s$. The sonic point temperature $T_s$ can be found by requiring that the right-hand side of Eq.~\eqref{eq:dvdr} be zero, using a simple root-finding algorithm. We then integrated $T$ and $\Phi$ outwards from the sonic point with Eq.~\eqref{eq:dTdr} and \eqref{eq:dphidr} to a maximum radius $r_\text{max}=10^9$ cm. We used these same equations to integrate inwards from the sonic point to $0.95\,r_s$, enough to step away from the sonic point and avoid numerical divergences. We then switched to integrating $r$ and $T$ with $\rho$ as the independent variable, all the way down to the surface of the star, constructing equations for $dr/d\rho$ and  $dT/d\rho$ from Eq.~\eqref{eq:drhodr} and Eq.\eqref{eq:dTdr}. 
We used $\rho$ as the independent variable instead of $r$ at this stage to avoid taking many small steps in $r$ in the geometrically thin region near the stellar surface, whereas $\rho$ changes by orders of magnitude in this region.

In order to have a single wind model per value of $\Mdot$, two boundary conditions must be imposed, which fix the final values $\Edot$ and $r_s$. Our inner boundary serves as a matching point between the wind and the burning layer.  Since the inner part of the wind close to the surface is in near-hydrostatic equilibrium, this relates to a pressure condition via the relation $P=gy$, where $g=(GM/R^2)\zeta
^{-1}(R)$ is the surface gravity of the star with radius $R$, and $y$ is the integrated column depth of the wind. We define the wind base $r_b$ as the location where a column depth $y_b=10^8$ g cm$^{-2}$ is reached, and require that this be the radius of the star, that is
\begin{equation}\label{eq:innerbc}
    r_b\equiv r(P/g=y_b)=R \,.
\end{equation}
Note that while we integrate our models to this high column depth, only a column $y_\text{ej}\sim 10^6$--$10^7$ g cm$^{-2}$ can be ejected in the wind, due to the limited supply of nuclear fuel \citep{Weinberg2006}. But by going to high pressure, we simply calculate the gradual transition to a shell in hydrostatic balance. Similarly, \citet{Paczynski1986b} required $r=R$ to occur at constant $T$, and \citet{Joss1987} required a constant $\rho$, both of which yield similar results. However, matching with $y$ is more useful if these wind models are to be used to connect to simulations of the burning layer, as $y$ is the more convenient coordinate in a hydrostatic layer. Finally, note that the exact spatial location of the matching point is not important and has no impact on the outer regions of the wind.

Our outer boundary condition is different from what has typically been used in the literature. Whereas previous studies used a thermal outer boundary condition such as $T=\Teff$ where $\kappa\rho r\approx 1$--$3$ (see \citealt{Paczynski1986b} and \S\ref{subsec:photospheres}), or added a free parameter $v_\infty$ to integrate inwards from a large radius (e.g. \citealt{Quinn1985}), we simply require that the velocity be finite at infinity,
\begin{equation}
    v(r\rightarrow\infty)>0\,.
\end{equation}
With our ability to integrate in the optically thin regions, we have found that the equations of structure \eqref{eq:dvdr}--\eqref{eq:dTdr} are very stiff, in that their behaviour at large radii is strongly dependent on the exact values of the initial conditions. This was studied in detail by \citet{Turolla1986}, who cautioned against trying to fine-tune initial parameters to shoot out to the correct solution at radial infinity. An alternative is to match to a second solution integrated inwards from infinity, as \citet{Quinn1985}. Instead, we integrated our solutions outwards with a simple but consistent step-wise shooting method, which we explain in Appendix \ref{sec:parameter_spaces}. This always worked and naturally lead the temperature profiles to the optically thin limit (eq.~[\ref{eq:T_thin}]).

Our procedure for determining the values for the parameters ($r_s$,$\Edot$) at every model value $\Mdot$ is a simple root-finding method for the two boundary conditions. First, for a range of values for $\Edot$, we find the value of $r_s$ that allows numerical integration to $r_\text{max}$ without having the velocity diverge in either direction. Then, we integrate inwards from $r_s$ to the wind base $r_b$, and evaluate the error on the boundary condition Eq.~\eqref{eq:innerbc}. The final values of ($r_s$,$\Edot$) are then found by searching for where this error vanishes, i.e., where the inner boundary condition is satisfied. In Appendix \ref{sec:parameter_spaces}, we show a visualization of the ($\Mdot$,$\Edot$,$r_s$) parameter space and the root-finding procedure.

\subsection{Static expanded envelopes}\label{subsec:envelopes}
The equations of structure for relativistic static atmospheres can simply be taken from the wind equations with $v=0$. In particular, we may integrate the equations for $dT/dr$ and $d\rho/dr$, Eq.~\eqref{eq:dTdr}  and \eqref{eq:drhodr}. In the optically thick limit, these equations reduce to the same equations as \citet{Paczynski1986a}.

Because there is no mass loss, there is no net transfer of energy from the radiation to the gas, i.e., the luminosity must be conserved through the atmosphere. Therefore, each model, labelled by its photospheric radius $r\ph$, is parametrized by a unique value of $\Linf$, the luminosity seen by observers at infinity (in Eq.~\eqref{eq:Edot}, $\Linf=\Edot$). The local luminosity $L$ is then a function of the radial coordinate with
\begin{equation*}
    L(r)=\Linf \left(1-\frac{2GM}{c^2r}\right)^{-1}=\Linf \zeta(r)^{-2}.
\end{equation*}

The method to find envelope solutions is similar to the wind case. We start at some middle point where we can specify an initial condition, then integrate outwards to a maximum radius, and inwards to the stellar surface, and verify that boundary conditions are satisfied on both ends. We start with a trial value of $\Linf$ at the photosphere, $r=r\ph$. The photosphere is defined as the location where $T=\Teff$, the effective blackbody temperature of the atmosphere, as in \citet{Paczynski1986a}. This gives an initial condition for the temperature,
\begin{equation}\label{eq:env_Tph}
    T\ph=\left(\frac{L\ph}{4\pi r\ph^2 \sigma}\right)^{1/4},
\end{equation}
where $T\ph=T(r\ph)$, $L\ph=L(r\ph)$, and $\sigma=ac/4$. This condition fixes the FLD parameters of Eq.~\eqref{eq:FLD_lambda_x} to $x=0.25$, $\lambda\approx 0.309$, meaning the photosphere is neither optically thick nor thin, but in the transition between the two.

We then search for the value of the density at the photosphere, $\rho\ph=\rho(r\ph)$, that allows us to integrate outwards from the photosphere to a large radius $r_\text{max}$, and reach the expected optically thin limit given by Eq.~\eqref{eq:T_thin} -- this is our outer boundary condition. Similarly to the wind case, we found that this required high precision on the initial values, or the integration would diverge. In particular, we found that the \citet{Paczynski1986a} prescription for $\rho\ph$, based on fixing the optical depth of the photosphere (the outer boundary for their optically-thick calculations), always resulted in $\rho$ crashing to zero just after the photosphere. We will show in \S\ref{subsec:photospheres} that fixing the optical depth of the photosphere is not accurate for envelopes.

We finally integrate inwards to the surface, where we require the same condition based on column depth as in the wind case, Eq.~\eqref{eq:innerbc}. For each choice of $r\ph$, we search for the values of $(\Linf,\rho\ph)$ that lead to both outer and inner boundary conditions being satisfied. We show the $(\Linf,r\ph,\rho\ph)$ parameter space in the Appendix \ref{sec:parameter_spaces}. In this way, we construct a family of envelope solutions parametrized by $L^\infty$.

\begin{figure*}
    \centering
    \includegraphics{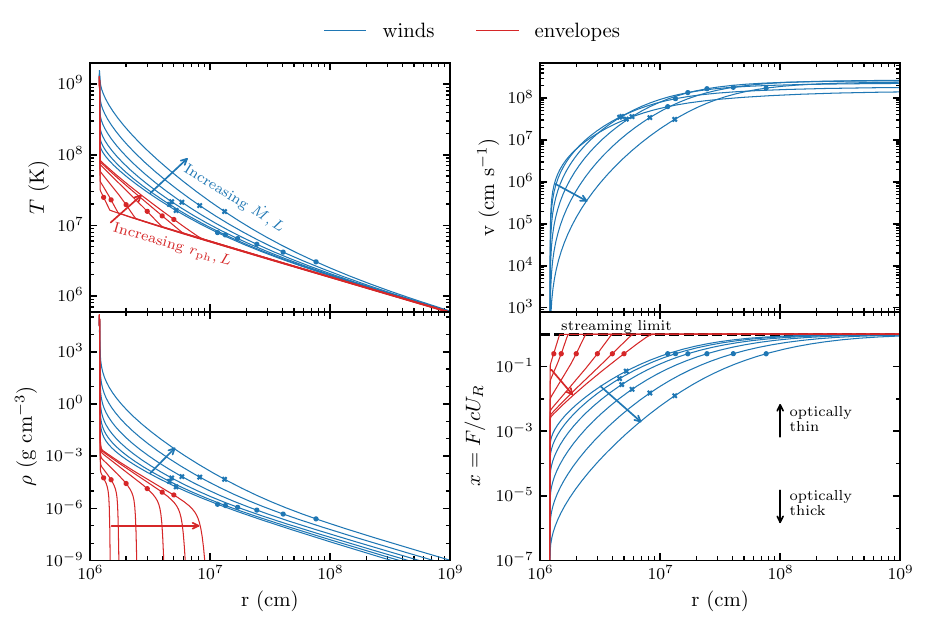}
    \caption{Radial profiles of the temperature (top left), velocity (top right), density (bottom left) and flux-energy ratio (bottom right), for pure helium winds (blue) and envelopes (red).  Crosses and dots indicate the positions of the sonic points and photospheric radii respectively.  Following the arrow direction, the winds have mass-loss rates $\log\dot{M}=(17.25,17.5,17.75,18.0,18.25,18.5)$ and base luminosities $L_b^\infty/\Ledd=(1.10,1.17,1.30,1.51,1.89,2.55)$. The envelopes have photospheric radii $r\ph=(13,15,20,30,40,50)$ km, and luminosities $\Linf/\Ledd=(0.89,0.92,0.95,0.98,0.99,1.00)$. The black arrows in the bottom right panel point towards the optically thick and thin limits.}
    \label{fig:profiles}
\end{figure*}

\begin{figure}
    \centering
    \includegraphics{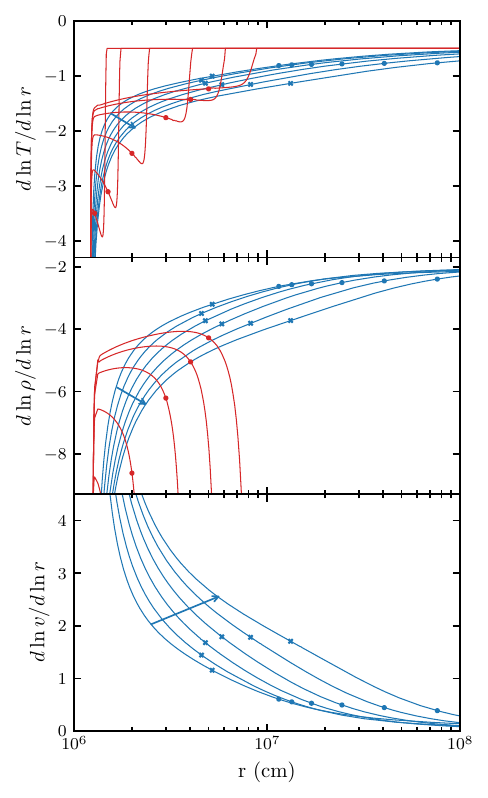}
    \caption{Gradients of the temperature, density and velocity for the same wind and envelope models as in Fig.~\ref{fig:profiles}. Crosses and dots mark the locations of $r_s$ and $r\ph$ respectively. Blue arrows show the direction of increasing $\Mdot$ for winds.}
    \label{fig:gradients}
\end{figure}

\section{Properties of the solutions}\label{sec:results}
We now discuss our solutions for winds and static envelopes. We calculate solutions for a neutron star of mass 1.4$M_\odot$ and radius 12 km, giving a surface gravity $g=1.60\times 10^{14}\ {\rm cm\ s^{-2}}$ and surface redshift factor $(1-2GM/Rc^2)^{-1/2}=1.24$. For the composition, we take fully ionized helium ($\mu=4/3$). We begin by showing profiles for temperature, density, velocity and flux in \S\ref{subsec:profiles}, comparing the static envelope and wind solutions.

 In \S\ref{subsec:photospheres}, we discuss our treatment of the photosphere and compare it to previous papers which did not model the optically thick to thin transition. In \S\ref{subsec:masslossrate}, we discuss the maximum mass-loss rate for winds, and in \S\ref{subsec:massloss}, we discuss the dependence of the mass-loss rate on base luminosity.

\subsection{Solution profiles}\label{subsec:profiles}
Fig.~\ref{fig:profiles} shows the radial profiles of temperature, density, velocity and flux for solutions of winds and envelopes. Close to the surface, both temperature and density drop sharply with radius in a thin layer in hydrostatic equilibrium (in the wind case the velocities are very subsonic at the stellar surface, so hydrostatic balance applies). In winds, the hydrostatic region slowly transitions to an outflowing one, where $v$ settles to ${\sim}\,0.01\,c\sim 3\times 10^8\ {\rm cm\ s^{-1}}$ at large distance.  Note that in the wind solutions, the profiles continue smoothly through the photosphere, whereas in envelopes the density drops exponentially above the photosphere. In all cases, at large distances the temperature goes to the correct optically thin power law, Eq.~\eqref{eq:T_thin}. For winds, the luminosity at large distance is within 1\% of $\Ledd$. The bottom right panel shows the radiative flux parametrized by $x$ (eq.~[\ref{eq:FLD_lambda_x}]). When $x\ll 1$, $\lambda\approx 1/3$ and the photons diffuse through the optically thick medium. The photosphere is at a fixed value of $x=0.25$ because of Eq.~\eqref{eq:env_Tph}, and shortly after the streaming limit $x=1$ is reached.

We show the gradient of $T$, $v$, and $\rho$ in Fig.~\ref{fig:gradients}. At the photosphere, the gradients are close to those expected for a constant velocity, in which case mass conservation (eq.~[\ref{eq:Mdot}]) implies $\rho\propto r^{-2}$ (actual values near the photosphere are $d\ln v/d\ln r\approx 0.5$ and $d\ln\rho/d\ln r\approx -2.5$, similar for all $\dot M$). Near the sonic point, these gradients are steeper, $d\ln v/d\ln r\approx 2$ and $d\ln\rho/d\ln r\approx -4$ for the larger $\dot M$ winds; $d\ln v/d\ln r\approx 1$ and $d\ln\rho/d\ln r\approx -3$ for the lowest $\dot M$. For comparison, \cite{Titarchuk1994} and \cite{Shaposhnikov2002} assumed that $v\propto r$ and $\rho\propto r^{-3}$ as an approximate background profile for their radiative transfer calculations. We see that in reality the power law indices vary continuously with $r$ through the region near the photosphere.

\begin{figure}
    \centering
    \includegraphics{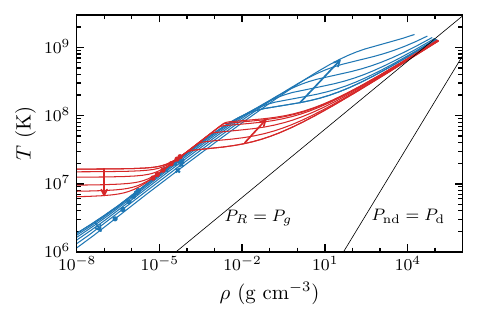}
    \caption{Density-temperature profiles of the same wind and envelope models as in Fig.~\ref{fig:profiles}. Crosses and dots mark the locations of $r_s$ and $r\ph$. Blue and red arrows show the direction of increasing $\Mdot$ and $r\ph$ for winds and envelopes respectively. The black lines indicate the points of transition from one pressure regime to another. $P_\text{nd}\propto T\rho$ and $P_\text{d}\propto\rho^{5/3}$ are the pressures of non-degenerate and degenerate electrons.}
    \label{fig:rho_T}
\end{figure}

Fig.~\ref{fig:rho_T} shows the corresponding density--temperature profiles. The models are all radiation pressure dominated in the extended regions, transitioning to a gas dominated regime in the surface regions. Note that at the high temperatures at the base of these envelopes, electron degeneracy is lifted so that our approximation of ideal gas applies even near the base.

\begin{figure}
    \centering
    \includegraphics{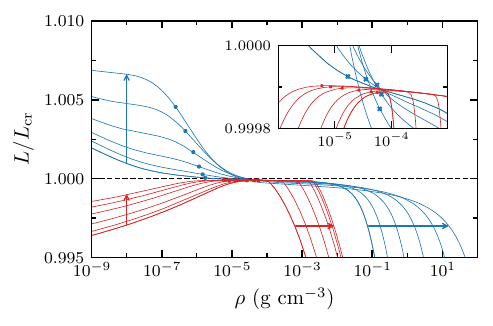}
    \caption{Density profiles of the luminosity in the same wind and envelope models as in Fig.~\ref{fig:profiles}. Crosses and dots mark the locations of $r_s$ and $r\ph$. Blue and red arrows show the direction of increasing $\Mdot$ and $r\ph$ for winds and envelopes respectively. The dashed black line denotes $L=\Lcr$. The inset zooms in to the locations of the wind sonic points and envelope photospheres.}
    \label{fig:luminosity}
\end{figure}

The sonic points for the winds range from ${\approx}\,40\ {\rm km}$ to more than $100\ {\rm km}$ (indicated by crosses in Fig.~\ref{fig:profiles}). It is interesting to note that the sonic points of these super-Eddington winds are much closer to the star than in a thermally-driven wind. The standard isothermal wind has $r_s=GM/2c_s^2\sim 10^6\ {\rm km}\ (T/10^7\ {\rm K})^{-1}$ (e.g.~\citealt{Parker1963}). In our case, setting the right-hand-side of Eq.~\eqref{eq:dvdr} (the numerator of $dv/dr$) equal to zero, and assuming that $\lambda\approx 1/3$ and $c_s^2\ll c^2$ at $r=r_s$, gives  
\begin{equation}\label{eq:rs_analytic}
    r_s \approx \frac{GM}{2c_s^2}\zeta^{-2}\left[ 1 - \left(\frac{4-3\beta}{ 4-4\beta}\right)\frac{L}{\Lcr} \right].
\end{equation}
With a sonic point temperature of $\approx 2\times 10^7\ {\rm K}$ and density $\approx 10^{-4}\ {\rm g\ cm^{-3}}$, $\beta \sim 10^{-4}$. Similarly, the luminosity lies slightly below the critical luminosity at the sonic point by a similar amount, $1-L/\Lcr\sim 10^{-4}$, as we show in Fig.~\ref{fig:luminosity}. The overall effect is to reduce the sonic point radius to $\sim 10^{-4} GM/2c_s^2\sim 100\ {\rm km}$. 

\subsection{Optical depth at the photosphere}\label{subsec:photospheres}

Optically-thick calculations typically set their outer boundary by specifying the optical depth at the photosphere, where $T=\Teff$. For example, for their static envelope models, \citet{Paczynski1986a} set the optical depth of the photosphere to be $\tau=2/3$ (corresponding to the plane-parallel grey atmosphere). Specifying the optical depth in turn specifies the pressure through the relation $P=gy=g\tau/\kappa$. In wind calculations, typically the optical depth parameter $\tau^\star = \kappa\rho r$ is specified at the photosphere, e.g. \citet{Paczynski1986b} set $\tau^\star=3$ when $T=\Teff$ and \cite{Herrera2020} set $\tau^\star=8/3$. 

\begin{figure}
    \centering
    \includegraphics{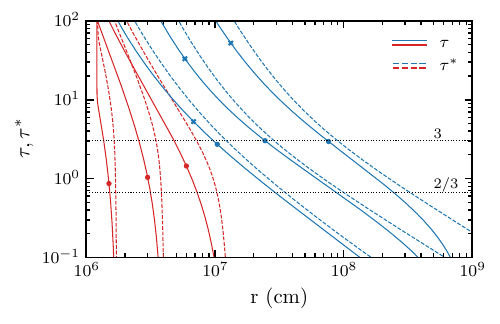}
    \caption{Optical depth given by $\tau^*\equiv\rho\kappa r$ and $\tau$ (eq.~[\ref{eq:tau}]) throughout the wind (blue) and envelope (red) models. The wind models shown are log$\Mdot=(17,18,18.5$, left to right), the envelope models are $r\ph=(15,30,60)$ km. Crosses and dots mark the locations of $r_s$ and $r\ph$. The dotted lines mark specific values of 2/3 and 3.}
    \label{fig:optical_depth}
\end{figure}

Figure \ref{fig:optical_depth} compares these optical depths at the photosphere (where $T=\Teff$) for our solutions. Since we model optically thin regions and go to arbitrarily large radii, we can calculate the true optical depth,
\begin{equation}\label{eq:tau}
    \tau(r)=\int_r^\infty \kappa(r)\rho(r) \frac{dr'}{\zeta(r')},
\end{equation}
where the curvature parameter $\zeta$ is included to give the proper length in the Schwarzschild geometry \citep{Niedzwiecki2006}. We see that for winds, $\tau$ and $\tau^\star$ are approximately equal to each other. Indeed, one can show that $\tau\approx\tau^*/(n-1)$ if $\rho\sim r^{-n}$ \citep{Quinn1985}, and as discussed in \S\ref{subsec:profiles}, $\rho$ is a power law in $r$ with $n\approx 2$ at $r>r\ph$.

Figure \ref{fig:optical_depth} shows that the optical depth at the photosphere increases from $\tau<1$ for the most compact extended atmosphere solutions, to $\tau\approx 3$ for the winds. Whereas \citet{Paczynski1986a} assumed that $\tau=2/3$ at the photosphere of their static envelopes, we see that in fact $\tau$ increases as the envelopes become more extended. For the winds, $\tau(r\ph)\approx$ 2.7--3, and $\tau^*(r\ph)\approx$ 3.7--3.9 which compare well with the values $\tau^\star=3$ and $5$ used by \citet{Paczynski1986b}. \cite{YuWeinberg2018} also found $\tau\approx 3$ for the photosphere in their (Newtonian) hydrodynamic calculations.
The result that $\tau=3$ for the wind solutions matches the expectation for a grey spherical atmosphere with a constant opacity and density profile $\rho\propto r^{-2}$. \cite{Larson1969} derived an expression for the temperature profile in the case where $\rho\kappa\propto r^{-n}$, finding
\begin{equation}
    T^4 = \frac{L}{16\pi r^2\sigma}\left[1 + 3\tau \left(\frac{n-1}{n+1}\right)\right].
\end{equation}
For $n=2$, this gives $T^4 = (\Teff^4/4)(1 + \tau)$, which implies that $T=\Teff$ for $\tau=3$. When $\tau\rightarrow 0$, we recover the optically thin limit Eq.~\eqref{eq:T_thin}.

\subsection{The maximum mass-loss rate}\label{subsec:masslossrate}

Figure \ref{fig:rho_T} shows that as $\Mdot$ increases, the transition from radiation pressure to gas pressure being dominant happens at larger and larger densities. The highest $\Mdot$ models are still radiation dominated ($\beta\ll 1$) at the base $r_b$. \cite{Paczynski1986b} pointed out that because models with high $\Mdot$ maintain the steeply increasing temperature to large depths, they are too hot to match onto a nuclear burning envelope, placing an upper limit on the mass-loss rate that can be achieved.

We can relate the expected mass-loss rate to the amount of nuclear energy deposited in the burst. In Appendix \ref{sec:base_conditions}, we derive an analytic expression for the value of $\beta$ expected in the burning layer after a release of nuclear energy $E_{\rm nuc} = E_{18} 10^{18}\ {\rm erg\ g^{-1}}$. In the limit of large energy release, we find $\beta\approx 4k_BT_R/\mu m_p E_{\rm nuc}\approx 0.06\,g_{14}^{1/4}y_8^{1/4}\mu^{-1}E_{18}^{-1}$, where $T_R$ is the radiation-pressure-limited temperature given by $gy = aT_R^4/3$. Burning helium to carbon releases $E_{\rm nuc}\approx 0.6\ {\rm MeV}$ per nucleon, and complete burning of helium to iron group elements gives $E_{\rm nuc}\approx 1.6\ {\rm MeV}$ per nucleon. So we expect the amount of energy that can be released rapidly at the start of a burst to be $E_{18}\sim 1$, meaning that the value of $\beta$ achieved in the initial stages of a burst is limited to be $\gtrsim 0.1$, with a corresponding limit on the mass-loss rate. 

\begin{figure}
    \centering
    \includegraphics{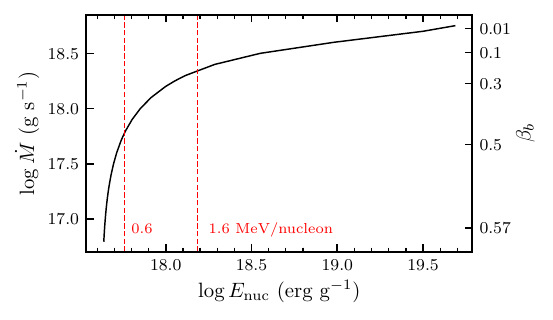}
    \caption{Values of the mass-loss rate $\Mdot$ as a function of the nuclear energy produced at the base, according to Eq.~\eqref{eq:Enuc2}. The right axis shows values of $\beta$ at the wind base. The dashed red lines mark the energy released by complete burning of helium to carbon and complete burning of helium to iron group elements.}
    \label{fig:base_beta}
\end{figure}

Figure \ref{fig:base_beta} shows the mass-loss rate $\dot M$ as a function of the energy deposited $E_{\rm nuc}$. We calculate this curve using our wind models to relate $\beta$ and $\dot M$, and using the result in Appendix \ref{sec:base_conditions} to relate $\beta$ to $E_{\rm nuc}$. We see that a nuclear energy release of order $1\ {\rm MeV}$ per nucleon limits the mass-loss rate in the resulting wind to be $\lesssim 2\times 10^{18}\ {\rm g\ s^{-1}}$. A similar sharp increase in the enthalpy per particle with mass-loss rate can be seen in Table 2 of \cite{Kato1983a}. In what follows, we show results only for wind models going up to $\Mdot=10^{18.5}$ g s$^{-1}$, corresponding to $L_b^\infty\approx 2.5\,\Ledd$. Note that \citet{Paczynski1986b} rejected models with $\Mdot>10^{19}$ g s$^{-1}$ for similar reasons (their larger $\Mdot$ range was because they matched to a higher temperature at the base). \citet{Herrera2020} found no solutions with $\Mdot\gtrsim 10^{19.5}$ g s$^{-1}$, even with a more flexible inner boundary condition. 

Figure \ref{fig:base_beta} also shows that the mass-loss rate drops off dramatically for $E_{\rm nuc}\lesssim 0.4\ {\rm MeV}$ per nucleon. This is roughly consistent with the estimate from \cite{Fujimoto1987} that in order for a burst to show radius expansion, the helium fraction in the fuel layer at ignition should be $\gtrsim 0.5$. The implication is that pure helium flashes should be able to readily provide enough nuclear energy to drive a wind. Expanded envelopes have similar values of $\beta$ at the base to the low $\dot M$ wind solutions ($\beta_b\approx$ $0.59$--$0.63$ for envelopes shown in Fig.~\ref{fig:profiles}--\ref{fig:optical_depth}), and so require similar energy releases $\approx 0.4\ {\rm MeV}$ per nucleon.

\subsection{Analytic formula for the mass-loss rate}\label{subsec:massloss}

\begin{figure}
    \centering
    \includegraphics{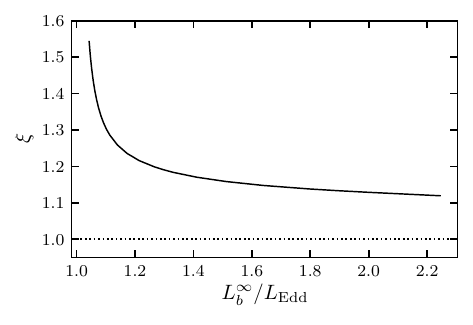}
    \caption{Ratio of predicted mass-loss rates from Eq.\eqref{eq:mdotprescription} to our model values, $\xi\equiv(L_b^\infty-\Ledd)/(GM\Mdot/R)$, as a function of the base luminosity redshifted to infinity.}
    \label{fig:Mdotprescription}
\end{figure}

A common prescription for mass-loss rates is the one derived by \citet{Paczynski1986b} and given in Eq.~\eqref{eq:mdotprescription}. This is obtained by equating the constant $\dot E$ (eq.~[\ref{eq:Edot}]) at the base of the envelope with that at the photosphere, neglecting the contributions from the enthalpy and kinetic energies, which are sub-dominant. In Fig.~\ref{fig:Mdotprescription}, we plot $\xi\equiv(L_b^\infty-\Ledd)/(GM\Mdot/R)$, the ratio of the prediction to our model values, as a function of the base luminosity. At low base fluxes and mass-loss rates, the prediction is 50\% larger than the true value, and remains 10\% larger even at higher mass-loss rates. 

These differences are straightforward to understand. At the base of the envelope, $\dot E\approx L_b^\infty + \dot M c^2 (1-2GM/Rc^2)^{1/2}$. At large distance from the star, $\dot E\approx \Ledd + \dot M (c^2 + w)$, where $w$ is the enthalpy per unit mass. We neglect the contribution from enthalpy at the base, assume the gravitational redshift factor is unity far away from the star, and neglect the kinetic energy $\propto v^2$ at both locations. With Eq.~\eqref{eq:T_thin} for $T$ in the optical thin region, and writing $\rho = \dot M/4\pi r^2 v_\infty$, the enthalpy in the outer part of the wind, $w=2aT^4/\rho$, can be written
\begin{equation}\label{eq:enthalpy_infinity}
    \frac{\dot M w}{\Ledd} = \frac{2v_\infty}{c},
\end{equation}
independent of $r$. Equating the two expressions for $\dot E$ and solving for the mass-loss rate gives 
\begin{eqnarray}\label{eq:Mdot_anal}
\dot M c^2\left[1-\zeta(R)\right]= L_b^\infty-\Ledd\left(1+\frac{2v_\infty}{c}\right)
\end{eqnarray}
where $\zeta(R)=(1-2GM/Rc^2)^{1/2}$ is the curvature parameter at the base. With $c^2(1-\zeta(R))\approx GM/R$ and neglecting the $v_\infty$ term, this reduces to Eq.~\eqref{eq:mdotprescription}. Without making these approximations, we find
\begin{eqnarray}\label{eq:xi_analytic}
\xi &=& \frac{L_b^\infty-\Ledd}{GM\Mdot/R}\nonumber\\&=&\frac{1-\zeta(R)}{ GM/Rc^2}\left[1-\frac{2v_\infty}{c}\frac{\Ledd}{L_b^\infty-\Ledd}\right]^{-1}.
\end{eqnarray}
The first term in equation \eqref{eq:xi_analytic} is ${\approx}\,1.12$ for our choice of $M$ and $R$, and causes the overall offset of $\xi$ from the dotted line in Fig.~\ref{fig:Mdotprescription}. The second term causes $\xi$ to increase sharply at low luminosities, where the enthalpy at large distances is significant, and consequently Eq.~\eqref{eq:mdotprescription} overpredicts the true mass-loss rate.

\section{The transition between expanded envelopes and winds}\label{sec:transition}

In this section, we discuss the evolution of the envelope during PRE bursts, where the luminosity $L_b^\infty$ is time-dependent. We explore our grid of models as a function of the base flux, and discuss its applicability for quasi-static calculations of bursts. 

\subsection{Comparison between expanded envelopes and winds as a function of base luminosity}\label{subsec:transition1}

\begin{figure}
    \centering
    \includegraphics{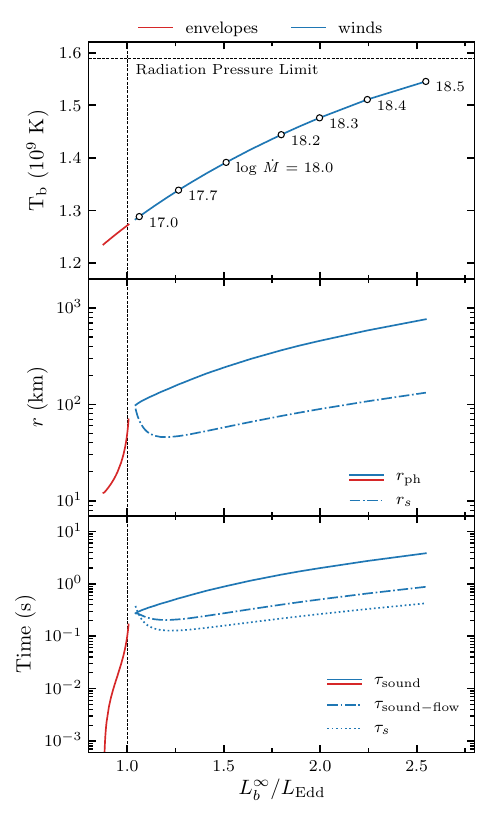}
    \caption{Solution space for the base luminosity as seen by observers at infinity, for both winds (blue) and envelopes (red).  Eddington luminosity is marked by the vertical black dashed line. \textit{Top}: Temperature at the base.  Mass-loss rates are indicated at various points for the winds.  \textit{Middle}: Sonic and photospheric radii of solutions.  \textit{Bottom}: Characteristic timescales of the solutions (see \S\ref{subsec:transition}).}
    \label{fig:triple}
\end{figure}

Figure~\ref{fig:triple} shows the base temperature (at a column depth $y_b=10^8\ {\rm g\ cm^2}$), photospheric and sonic radius, and sound-crossing and flow timescales, all as a function of the base luminosity in Eddington units. The static envelope models lie to the left ($L_b^\infty\lesssim \Ledd$) and the winds to the right ($L_b^\infty\gtrsim \Ledd$). The corresponding values of $\dot M$ are shown in the top panel. 

In the top panel of Fig.~\ref{fig:triple}, we see that the envelope and wind models lie on a common track in the $T_b$--$L_b$ plane. The scaling is close to $L_b\propto T_b^4$. At large $\dot M$, the base temperature approaches the radiation-pressure-limiting temperature given by $gy = aT_R^4/3$, or $T_R = 1.59\times 10^9\ {\rm K}\ (g_{14}/1.6)^{1/4}y_8^{1/4}$ (see Appendix \ref{sec:base_conditions}). As discussed in \S\ref{subsec:masslossrate}, we show models up to a maximum mass-loss rate of $3\times 10^{18}\ {\rm g\ s^{-1}}$. The static envelope models extend slightly to the right of the $L_b^\infty/\Ledd=1$ line; this is because there is a small Klein-Nishina correction to the electron scattering opacity in the outer layers of our models where $T\sim 10^7\ {\rm K}$ (eq.~[\ref{eq:kappa}] gives $\kappa_0/\kappa\approx 1.04$ at $T=10^7\ {\rm K}$).

We show wind solutions down to a mass-loss rate of $\dot M = 10^{17}\ {\rm g\ s^{-1}}$. We have not been able to obtain solutions for lower mass-loss rates than this due to numerical difficulties (see Appendix \ref{sec:parameter_spaces} for a discussion); there is a small range of luminosity between $1\lesssim L_b^\infty/\Ledd\lesssim 1.05$ where we do not have a solution. As shown in the middle panel of Fig.~\ref{fig:triple}, this value of $\dot M$ corresponds to the point at which the critical point of the wind becomes optically thin ($r_{\rm ph}\approx r_s$).  As discussed by \cite{Kato1994} in the context of nova winds, acceleration becomes very inefficient when the sonic point is far outside the photosphere; steady-state wind solutions may then not be possible. \cite{Kato1994} argued that the envelopes should smoothly evolve into wind solutions, and adjusted the boundary condition in their optically-thick models so that was the case. However, we do not see this behaviour in our models, but instead have a small luminosity difference between the maximally-extended static atmosphere and the lowest $\dot M$ optically-thick wind.

With the exception of the gap between $1\lesssim L_b^\infty/\Ledd\lesssim 1.05$, the middle panel of Fig.~\ref{fig:triple} shows that the photospheric radius in our models does appear to smoothly evolve from the static envelopes to the winds. The maximally extended envelope that we find has $r_{\rm ph}\approx 70\ {\rm km}$, whereas the lowest $\dot M$ wind has $r_{\rm ph}\approx 100\ {\rm km}$. Note that this is very different in the optically thick models. \cite{Paczynski1986a} found static envelopes with photospheric radius $\approx 200\ {\rm km}$, and we have found, reproducing their calculations, that $r_{\rm ph}\sim 1000\ {\rm km}$ is possible with the boundary condition $\tau=2/3$ at the photosphere. These large photospheric radii are much greater than the $\sim 100\ {\rm km}$ of the wind solutions. Our models show instead that the optical depth at the photosphere in the extended envelopes increases towards $\tau \approx 3$ as the envelope becomes more extended (see Fig.~\ref{fig:optical_depth}), and the photospheric radius monotonically increases, smoothly transitioning into the wind solutions.

\subsection{The timescale to reach steady-state}\label{subsec:transition}

With wind velocities of $v_\infty\approx 3\times 10^8\ {\rm cm\ s^{-1}}$, the flow timescale is $t_{\rm flow}=r/v_\infty\sim 0.03\ {\rm s}\ (r/100\ {\rm km})$. This is much shorter than the typical $\gtrsim 1\ {\rm s}$ timescale of the super-Eddington phases of bursts, justifying the use of steady-state hydrodynamic equations to calculate the wind structure. Indeed, as long as the ratio of flow time to evolution time remains small, it should be appropriate to use steady-state wind solutions as outer boundary conditions for calculations of the interior evolution \citep{Joss1987}. Because the wind photospheres with GR effects included remain at large radii $\sim 100\ {\rm km}$ even for small mass-loss rates, it means that the photosphere has to adjust by a large amount when the base luminosity crosses the Eddington limit, perhaps calling into question whether this quasi-static approach is applicable. 

The bottom panel of Fig.~\ref{fig:triple} shows the flow and sound crossing timescales for the different models. Hydrostatic balance is established on a timescale given by the sound crossing time,
\begin{equation}
    \tau_\text{sound}=\int_R^{r\ph}c_s^{-1}dr\,,
\end{equation}
which gives the time taken for a sound wave to travel the structure, from the base to the photosphere. For winds, a characteristic timescale for the flow is
\begin{equation}
    \tau_s=\frac{r_s}{c_s(r_s)}=\frac{r_s}{v(r_s)}\,.
\end{equation}
We could also define the flow crossing time
\begin{equation}\label{eq:flowtime}
    \tau_\text{flow}=\int_R^{r\ph}v^{-1}dr\, ,
\end{equation}
but this has the problem that the velocity is so small near the neutron star surface that the flow time is dominated by these regions, and is therefore not representative of the whole solution. Instead, we take a timescale that combines the sound crossing time of hydrostatic regions, up to the sonic point, followed by the flow crossing time in the outflowing regions of the wind, up to the photosphere,
\begin{equation}
    \tau_\text{sound-flow}=\int_R^{r_s}c_s^{-1}dr + \int_{r_s}^{r\ph}v^{-1}dr \,.
\end{equation}
Fig.~\ref{fig:triple} shows that these wind timescales have similar values and progressions with $L_b^\infty$, except for low mass-loss rates where the increase in sonic point radii results in larger crossing times.  In every model, by looking at the bottom two panels in Fig.~\ref{fig:triple}, it is clear that it is the photospheric radius which largely dictates the timescales. This means that more extended structures take longer to form, and that they cannot exist under a rapidly varying luminosity. 

Typical PRE bursts have a duration of $\lesssim 10\ {\rm s}$, with Eddington phases shorter than a few seconds \citep{Galloway2008}, although as discussed in \S\ref{subsec:observational_motivation}, there are bursts with much longer durations \citep{IntZand2010}. Since the rising phase is so fast in transitioning from sub to super-Eddington luminosities, it is clear that our stationary solutions are not appropriate for describing its dynamics. At the high end of luminosities for static envelopes, the extended region takes ${\sim}\,0.1$ s to adjust to a small change in surface flux and expand its photosphere further. This is too long for the burst rise to go through each solution in succession. A time-dependent calculation is therefore needed to model this part of the burst.  However, our timescales would allow for the Eddington and decaying phase to be reasonably be modelled by steady-state winds and envelopes respectively. The quasi-static approach is therefore appropriate to PRE model bursts once they reach Eddington. For bursts with $r\ph\ll100$ km, static envelope models can be used as well, since they never reach Eddington and cross into the wind regime. Our timescales appear consistent with the time-dependent Newtonian calculations of \cite{YuWeinberg2018}, which have fairly constant $\dot M$ profiles after $\approx 1\ {\rm s}$.

\begin{figure}
    \centering
    \includegraphics{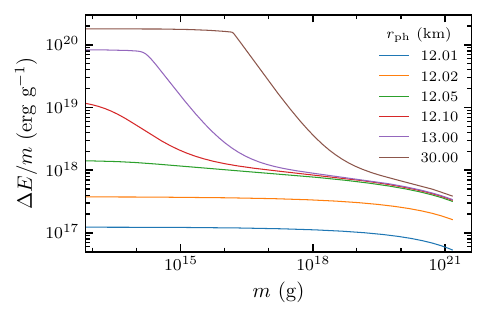}
    \caption{The energy per unit mass as a function of mass coordinate compared to an envelope with $r_{\rm ph}=12.005\ {\rm km}$. This shows the energy that must be given to the envelope above each mass coordinate in order to expand outwards. The mass coordinate $m$ measures the mass contained above a given point (to the photosphere). The total mass of the envelope is $\approx 10^{21}\ {\rm g}$.}
    \label{fig:env_energy}
\end{figure}

\subsection{Energetics of the expansion}

A further consideration in modelling the evolution of the envelope quasi-statically is the energy required to establish the expanded configuration, i.e.~to move from one model to the next as $L_b^\infty$ increases. We compute the energy of a given model above radius $r$ as
\begin{equation}
E(r)=\int_r^{r\ph}\left[U\zeta^{-1}+\rho c^2(1-\zeta^{-1})\right]4\pi r'^2dr',
\end{equation}
where the first term is the internal energy and the second is the gravitational binding energy \citep{Fowler1964}. The energy of the static envelope solutions is shown in Figure~\ref{fig:env_energy} as a function of the mass coordinate 
\begin{equation}\label{eq:env_mass}
    m\equiv \int_r^{r\ph}\rho\,4\pi r'^2\,dr',
\end{equation}
which measures the mass between radius $r$ and the photosphere. In order to focus on the change in energy as the envelope expands outwards, we show the difference in energy $\Delta E=E-E_0$, where $E_0$ is the energy of the most compact envelope that we calculated. 

We see that small increases in the photospheric radius (up to $\sim 50\ {\rm m}$) are associated with an energy per gram ${\lesssim}\,10^{18}\ {\rm erg\ g^{-1}}$ or equivalently ${\lesssim}\,1\ {\rm MeV}$ per nucleon that is fairly uniformly distributed across the envelope. With a total mass ${\sim}\,10^{21}\ {\rm g}$, the total energy is ${\sim}\,10^{39}\ {\rm erg}$ (note that this is roughly what we would expect to lift $m=10^{21}\ {\rm g}$ a height $H\sim 100\ {\rm m}$ with $g\sim 10^{14}\ {\rm cm\ s^{-2}}$, i.e.~$mgH\sim 10^{39}\ {\rm erg}$). At this point, the base temperature has increased to a value where radiation pressure is becoming important at the base of the model. Further expansion occurs by lifting the outer parts of the envelope outwards. For example, in the model with photospheric radius of $30\ {\rm km}$, the outermost ${\approx}\,10^{16}\ {\rm g}$ is given an energy ${\sim}\,10^{20}\ {\rm erg\ g^{-1}}\sim GM/R$, allowing it to  move to large radius. However, this represents only ${\sim}\,10^{-5}$ of the mass of the envelope; the rest remains in a compact configuration. 

Because only a small fraction of the mass is lifted outwards, the amount of energy required for the solutions with extended photospheres is not significantly more than the compact envelopes, and is readily supplied by nuclear burning. For example, in the model with $r_{\rm \ph}=30\ {\rm km}$, lifting $10^{16}\ {\rm g}$ outwards requires ${\sim}\,10^{36}\ {\rm erg}$, which is supplied in $<0.01\ {\rm s}$ at the Eddington luminosity. As long as the base luminosity evolves on a longer timescale, these energetic requirements are easily met.

For wind models, since the energy released from nuclear burning $E_\text{nuc}\sim 1$ MeV/nucleon is ${\sim}\,100$ times smaller than the gravitational binding energy $GM/R$, no more than ${\sim}\,1\%$ of the accreted mass may be fully ejected. In steady-state, this leads to the requirement that the ratio of mass within the sonic point to mass outside the sonic point $m_{r<r_s}/m_{r>r_s}$ should be larger than $\approx 100$ (where the mass $m$ is given by Eq.~\eqref{eq:env_mass} with modified integration bounds) \citep{Paczynski1986b}. Note that with the equation for the mass-loss rate (eq.~[\ref{eq:Mdot}]), this mass ratio can be re-written as the ratio of flow times (eq.~[\ref{eq:flowtime}]) between subsonic and supersonic regions (up to a factor of $\Psi$), which \citet{Quinn1985} used to rule out models with large $\Mdot$. In our case, these ratios are larger than 500 for all models. Our criterion for the maximum mass-loss rate based on $\beta_b$ (\S \ref{subsec:masslossrate}) turned out to be more restrictive.

\begin{figure*}
    \centering
    \includegraphics{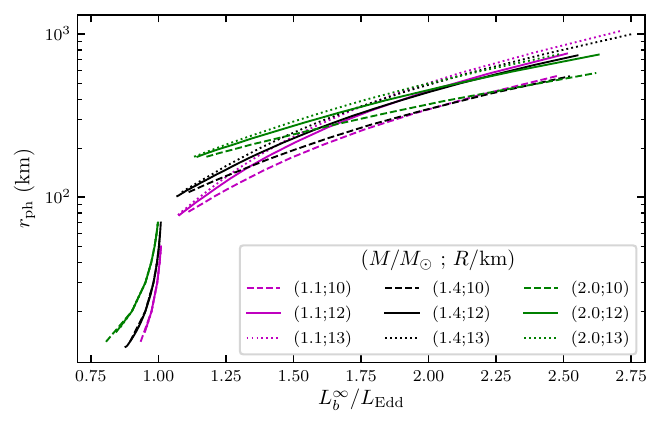}
    \caption{Photospheric radii of wind and envelope models for neutron stars with varying masses and radii. The solid black line represents the models shown throughout this paper.}
    \label{fig:spot_check_MR}
\end{figure*}

\section{Observational Implications}\label{sec:implications}

We now discuss the implications of our models for observations of PRE bursts. We first discuss the expected photospheric radii in PRE bursts (\S \ref{subsec:rph}). In \S\ref{subsec:spectral_shifts}, we calculate gravitational redshifts and velocity blueshifts for our wind models and review them in the context of \citet{Strohmayer2019}. In \S\ref{subsec:touchdown}, we discuss the location of the photosphere when the luminosity is close to Eddington and implications for identifying the touchdown point in observations of PRE bursts.

\subsection{Photospheric radii}\label{subsec:rph}

We find in our models that there is a separation between the photospheric radii of envelopes and winds. Static envelopes have photospheric radii ${\lesssim}\,100\ {\rm km}$; winds have $r_{\rm ph}>100\ {\rm km}$ (see middle panel of Figure~\ref{fig:triple}). Whereas optically-thick envelopes are able to extend to $r_{\rm ph}>100\ {\rm km}$, we find that once we relax that assumption, the photospheric radius evolves smoothly between envelopes and winds, as discussed in \S\ref{subsec:transition}.

Figure~\ref{fig:spot_check_MR} shows further results for the photospheric radius, now for different choices of neutron star mass and radius. Interestingly, for static envelopes and winds with low mass-loss rates, the photospheric radius is independent of neutron star radius, whereas for high mass-loss rate winds the photospheric radius becomes independent of neutron star mass. We also calculated pure hydrogen and solar composition ($X=0.7, Y=0.28, Z=0.02$) models, instead of pure He, and found only minor differences. The largest change in photospheric radius was an increase by $\approx 30$\% for H or solar composition winds compared to helium winds.

We define the photospheric radius as the location where the gas temperature $T$ is equal to the effective temperature $T_{\rm eff}$.
Observationally, the photospheric radius is inferred from blackbody fits, with a color correction applied to correct the measured color (blackbody) temperature to the effective temperature. 
In a scattering atmosphere with coherent scattering, the radiation temperature is set at the thermalization depth where the last absorption occurs and then photon scatter outwards to the scattering photosphere. \cite{Joss1987} found that the ratio of absorption to scattering opacity at the photosphere is ${\sim}\,0.01$, leading to thermalization depths $\approx 3$--$5$ times smaller than the scattering photosphere. However, they also showed that Compton scattering in the region between the thermalization depth and the scattering photosphere is effective in coupling the radiation and gas temperatures (although with constant photon number). 

For compact atmospheres near the Eddington limit, calculations of the spectrum find color correction factors $f_c=T_c/T_{\rm eff}\approx 1.5$--$2$ near Eddington luminosity \citep{Pavlov1991,Suleimanov2011}. The spectrum is harder than a blackbody, so that the blackbody radius underestimates the true emission radius by a factor of $f_c^2\approx 2$--$4$. For winds, \cite{Titarchuk1994} and \cite{Shaposhnikov2002} found that the situation reverses, $f_c<1$, so that the blackbody radius overestimates the true emission radius.

Observationally, most PRE bursts have modest expansions of tens of km \citep{Galloway2008}, consistent with being due to expanded atmospheres rather than winds. \cite{Paczynski1986b} made a similar point, that with large photospheres and therefore correspondingly low temperatures, wind solutions did not appear to match observations of bursts. Superexpansion bursts \citep{IntZand2010} do have blackbody radii of hundreds of km, consistent with wind solutions. The color corrections described earlier do not appear to change the conclusion that in the context of light element models, most observed PRE bursts do not have a wind.

\begin{figure}
    \centering
    \includegraphics{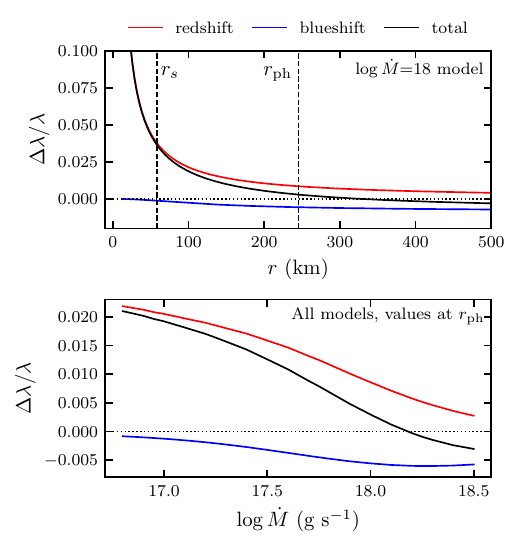}
    \caption{Wind spectral shifts, as a function of the radial coordinate for the log$\Mdot$ = 18 model (top), and as a function of $\Mdot$ at the photosphere (bottom).}
    \label{fig:lineshift}
\end{figure}

\subsection{Spectral shifts}\label{subsec:spectral_shifts}
Next, we investigate the importance of spectral shifts in our wind models. Redshift comes from relativistic curvature,
\begin{equation}
    1+\frac{\Delta\lambda_\text{red}}{\lambda_0}=\left(1-\frac{2GM}{r_0c^2}\right)^{-1/2}=\frac{1}{\zeta(r_0)}\,,
\end{equation}
where $r_0$ is the emission radius and $\lambda_0$ is the emission wavelength. Blueshift comes from the special relativistic Doppler effect,
\begin{equation}
    1+\frac{\Delta\lambda_\text{blue}}{\lambda_0}=\sqrt{\frac{1-v_0/c}{1+v_0/c}}\,,
\end{equation}
where $v_0$ is the gas velocity at $r_0$. In the top panel of Fig.~\ref{fig:lineshift}, we see that redshift dominates everywhere before the sonic point, where it can reach values of several percent. After the sonic point and approaching the photosphere, redshift and blueshift become comparable to the point where the total spectral shift is close to zero.  In the bottom panel, we note the changing sign of $\Delta\lambda$ at high $\Mdot$, as both velocities and photospheric radii increase, although being able observing such small shifts is unlikely.

We can see from the bottom panel of Fig.~\ref{fig:lineshift} the total shift of wind models at the photosphere is of at most 2\%. As for the relative shift of individual lines during different burst, if two lines $\lambda_1$ and $\lambda_2$ are shifted from their rest frame wavelength $\lambda_0$ by $\Delta\lambda_1$ and $\Delta\lambda_2$, then their relative shift is
\begin{equation}
    \frac{\lambda_1}{\lambda_2}=\frac{\lambda_0+\Delta\lambda_1}{\lambda_0+\Delta\lambda_2}\approx 1+\frac{\Delta\lambda_1}{\lambda_0}-\frac{\Delta\lambda_2}{\lambda_0}\,,
\end{equation}
so that relative shifts of at most ${\approx}\,1.02$ are expected. For a heavier star, e.g. with a $2\,M_\odot$ mass, the minimum photospheric radius is still $>100\ {\rm km}$ (Fig.~\ref{fig:spot_check_MR}), and the maximum relative shift increases by $1\%$. This only applies for winds, as static envelopes can have stronger redshifts given their smaller photospheric radii. \citet{Strohmayer2019} found a relative shift of ${\approx}\,1.046$, with the observed photospheres of the weaker bursts being at ${\approx}\,75$ km. Even so, gravitational redshift alone cannot explain this value (if the mass is ${\leq}\,2.0\,M_\odot$, see their Fig. 9), prompting them to suggest a blueshift contribution. Our results do not support this scenario, as the observed photospheric radii would imply static solutions rather than winds. Even in the wind case, models with small photospheric radii have very weak blueshifts (${\lesssim}\,1\%$) at the photosphere (Fig.~\ref{fig:lineshift}).

We must also note that the emission radius of spectral lines is likely not at the helium scattering photosphere. Heavier elements that are thought to be ejected have more complex interactions with radiation, and spectral lines and edges themselves are not a continuum effect, which is how radiation is treated in our model. Our wind models can describe the relative importances of redshift and blueshift, but true predictions on spectral lines will require a more sophisticated treatment of radiative transfer.

\subsection{Compact envelopes and touchdown radius}\label{subsec:touchdown}

We discussed in \S\ref{subsec:observational_motivation} the common technique of finding the neutron star radius based on measuring the touchdown flux, i.e., the flux when the temperature peaks and the photosphere presumably touches back down to the surface following the PRE phase of the burst. But if the luminosity at the touchdown point is still near-Eddington, an expanded envelope could be present, which means that the photospheric (touchdown) radius is not the neutron star radius. This difference enters into the redshift factor that relates the Eddington flux at the neutron star surface and the observed luminosity. To investigate this question, we have extended our calculation of envelopes to very compact ones with photospheres less than 1 km above the neutron star surface.

\begin{figure}
    \centering
    \includegraphics{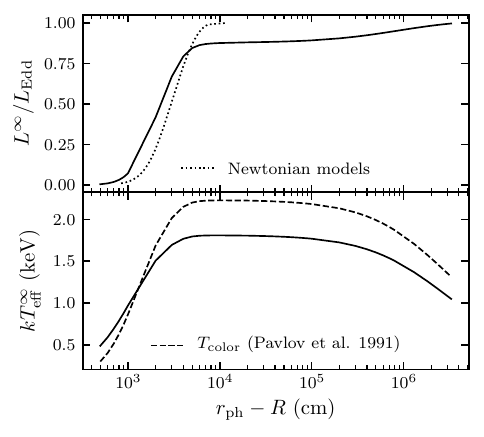}
    \caption{Difference between the photospheric (touchdown) radius based on the luminosity (top) and observed effective temperature (bottom). The dotted line in the top panel represents Newtonian atmospheres, for which we derive an analytical formula in Appendix \ref{sec:appendix_newtonian}. The dashed line in the bottom panel gives the color temperature of the atmosphere with the correction factor  of \citet{Pavlov1991}.}
    \label{fig:touchdown}
\end{figure}

The top panel of Fig.~\ref{fig:touchdown} shows the relation between the luminosity and the extension of the photosphere above the neutron star surface. The range of luminosities that would cause a significant difference between the touchdown and neutron star radius ($\gtrsim 100$ m) is quite narrow, from ${\sim}\,0.85\,\Ledd$ to ${\sim}\,\Ledd$. As \citet{Paczynski1986a} pointed out, this large expansion is only possible in GR, as Newtonian envelopes can only be compact; we show Newtonian models as a dotted line in Fig.~\ref{fig:touchdown} (an analytic solution for Newtonian envelopes is derived in Appendix \ref{sec:appendix_newtonian}). 
As luminosity increases, the envelope begins to expand significantly when the local luminosity at the surface of the star first becomes critical. Our range of luminosities where significant expansion occurs is smaller than \citet{Paczynski1986a}, who predicted expansion for $L\gtrsim 0.77\,\Ledd$, derived with $L\ph\approx \Lcr(r\ph)\approx \Ledd/\zeta(r\ph)$, assuming $\kappa(r\ph)\approx\kappa_0$. This second approximation is incorrect, since the temperature at the photosphere for models with moderate expansion can still be quite high. For example, the model with $r\ph=12.1$ km has a photospheric temperature of $2.6\times 10^7\ {\rm K}$, giving  $\kappa(r\ph)=0.92\,\kappa_0$ and thus $\Linf=0.87\,\Ledd$.

In the bottom panel of Fig.~\ref{fig:touchdown}, we plot the effective temperature $\Teff$ of the envelope (redshifted to infinity with $T^\infty=\zeta T$) as a function of the photospheric radius. To check for potential variations due to color corrections, we also show the expected color temperature calculated using the analytic result from \citet{Pavlov1991} that applies to compact atmospheres. The peak temperature occurs when the photosphere is close to the surface, within ${\sim}\,30\ {\rm m}$. However, the temperature is very flat with increasing $r\ph$, which means that a small uncertainty in the peak temperature can correspond to a relatively large uncertainty in the position of the photosphere. For example, when the photosphere is $1\ {\rm km}$ from the surface, the observed temperature is within 3\% of its peak value.

Our results provide evidence for the suggestion of \cite{Steiner2010} that the photosphere might still not have returned fully to the neutron star surface at the point of the burst identified as touchdown. 
They argued that the ratios of photospheric radius to neutron star radius implied by their analysis were $>1.1$, $>1.4$ and $>5$ for three different data sets that they considered. The last of these, a factor of 5, does not seem likely since the temperature is then well away from its peak value (even when including possible color correction variations; see \citealt{Ozel2016}). However, of order $\sim 10$\% overestimation of the photospheric radius at touchdown due to uncertainties in the measured temperature are more plausible given the results in Fig.~\ref{fig:touchdown}. Note that because the photospheric radius at touchdown appears inside a redshift factor, the corresponding uncertainty in the inferred neutron star radius will be smaller still by a factor of a few.

\section{Summary and discussion}\label{sec:discussion}

We constructed a sequence of self-consistent models of light element static expanded envelopes and steady-state winds resulting from near or super-Eddington luminosities in type I X-ray bursts\footnote{The models are available at \url{https://github.com/simonguichandut/GR-FLD-PRE}.}. We included general relativistic corrections that are necessary to correctly model the expansion of the envelope when the photosphere is close to the stellar surface. We also improve upon the earlier work of \cite{Paczynski1986a} and \cite{Paczynski1986b} by using flux-limited diffusion to model the transition from the optically-thick to optically-thin parts of the envelope. The optical depth at the photosphere then naturally transitions from close to $\tau=2/3$ for geometrically thin envelopes to $\tau\approx 3$ in the wind solutions (Fig.~\ref{fig:optical_depth}).

With this self-consistent treatment of the photosphere in hand, the models give the following picture for the evolution during the rising phase of a PRE burst. At low luminosity, the envelope is geometrically thin, and undergoes modest expansion with increasing luminosity. However, eventually, the luminosity at the stellar surface reaches the critical luminosity (the local Eddington limit), and the envelope adjusts by expanding outwards. Over a narrow range of luminosity from $0.8\lesssim L_b^\infty\lesssim 1$, the envelope remains in hydrostatic balance but the photospheric radius moves smoothly outwards to a radius of $\approx 50$--$80\,{\rm km}$ when $L_b^\infty\approx \Ledd$ (Fig.~\ref{fig:spot_check_MR}). For $L_b^\infty\gtrsim 1.05\,\Ledd$, an optically-thick wind develops, with photosphere located at $>100\ {\rm km}$. The radiative luminosity of the wind is within 1\% of the Eddington luminosity, with the remaining energy used to eject matter, thereby the base luminosity sets the mass-loss rate (eq.~[\ref{eq:Mdot_anal}]). The maximum mass-loss rate is $\approx 2\times 10^{18}\ {\rm g\ s^{-1}}$, determined by the available nuclear energy (Fig.~\ref{fig:base_beta}). 

We find that there are two aspects of the burst rise that are likely not well-modeled by quasi-static evolution. For the static expanded envelopes, the large photospheric radius means that the sound crossing time of the envelope can be $\sim 0.1\ {\rm s}$, comparable to burst rise times (Fig.~\ref{fig:triple}). Second, in the range $1\lesssim L_b^\infty/\Ledd\lesssim 1.05$, the sonic point of the wind is optically thin and we were not able to find steady-state solutions (\S\ref{subsec:transition1}). Time-dependent calculations are needed to model the evolution of the envelope as the base luminosity traverses the range from $\approx 0.8\,\Ledd$ to $\approx 1.05\,\Ledd$. \cite{YuWeinberg2018} recently carried out the first time-dependent calculations of PRE bursts, and the onset of a wind, but in Newtonian gravity. It would be extremely interesting to extend this kind of work by including general relativistic corrections.

While it is encouraging that the range of photospheric radii in models agrees well with the range of observed blackbody radii in PRE bursts ($\sim 10$--$1000\ {\rm km}$; Fig.~\ref{fig:spot_check_MR}), the fact that most observed PRE bursts show only modest expansion of a few tens of km is hard to explain. Photospheric radii $\lesssim 100\ {\rm km}$ can be achieved with static envelopes, but only for a narrow range of luminosities. This requires a fine-tuning of the energy release in the burst. A much more natural outcome of the models is that the luminosity at the base exceeds $\Ledd$, resulting in a wind with photospheric radius $\gtrsim 100\ {\rm km}$. 

As proposed by \cite{IntZand2010} and \cite{YuWeinberg2018}, heavy elements in the wind could truncate the wind and lead to smaller photospheres in agreement with observations. \cite{IntZand2010} pointed out that line-driving due to hydrogenic heavy ions could play an important role. This may be particularly important to include for luminosities just above Eddington where the sonic point is optically thin and radiation pressure driving is inefficient. A truncated wind might also more naturally explain the lineshifts observed by \cite{Strohmayer2019}, which are larger than predicted at the photospheres of our wind solutions. 

The expanded envelope models also have implications for neutron star radius measurements by the touchdown method. Our models show that the photosphere is still $1\ {\rm km}$ above the surface when the effective temperature is only 3\% away from its maximum value. This is a possible systematic uncertainty when interpreting the measured Eddington fluxes from bursts at touchdown. This uncertainty is not present when considering Newtonian envelopes, which remain within $\approx 100\ {\rm m}$ of the surface even close to the Eddington luminosity. The expanded nature of the envelope for $L^\infty\gtrsim 0.8\,\Ledd$ should be included in spectral models.

The models presented in this paper are based on many assumptions, namely a non-rotating neutron star, no magnetic fields, spherical symmetry and steady-state outflows. To take full advantage of the observational data of PRE bursts, more work is needed to drop each of these assumptions. In these accreting systems, the neutron star can have a short spin period such that the effective gravity at the surface is significantly reduced ($R\Omega^2/g=0.01g_{14}^{-1}(R/10\,\text{km})(f/500\,\text{Hz})^2$). This effect changes along the latitude so that the photosphere may lift off at different times during the burst rise, or be differently extended at touchdown \citep{Suleimanov2020}. Further, magnetic field lines may entrain the ionized fluid out to an \textit{Alfv\'en radius}, i.e. the point $r=r_A$ where $\rho u^2/2=B^2/8\pi$ \citep{LamersCassinelli}. For example, in a split monopole configuration for the magnetic field, $B(r)=B_0(R/r)^2$, and a surface value $B_0=10^9$ G, our wind models give $r_A\approx500$--5000 km, much farther out than the photosphere. For a dipole magnetic field $B\sim r^{-3}$, $r_A\approx 100$--300 km, closer in but still past the sonic point. This simple analysis shows that magnetic fields could very well have a strong influence on the dynamics of these outflows, and should be taken into account in future calculations. Lastly, the steady-state assumption prevents us from studying the evolution of multiple-stage bursts, such as ones in which the ejection of a hydrogen shell precedes the helium flash (see \citealt{Kato1986} for discussion), as is thought to have recently been observed by \textit{NICER} \citep{Bult2019}. For all of these reasons, future work on PRE bursts modeling should aim towards multidimensional radiation magnetohydrodynamics calculations.

\section*{Acknowledgements}
We thank Ninoy Rahman for useful discussions on flux-limited diffusion in general relativity. We also thank the anonymous referee for useful comments. SG is supported by an FRQNT scholarship. AC is supported by an NSERC Discovery grant and is a member of the Centre de Recherche en Astrophysique du Qu\'ebec (CRAQ). ZL and AC thank the International Space Science Institute in Bern for hospitality. ZL was supported by National Natural Science Foundation of China (U1938107, U1838111), and Scientific Research Fund of Hunan Provincial Education Department (18B059).

\bibliographystyle{aasjournal}
\bibliography{references}

\appendix

\section{Derivation of the steady-state hydrodynamics equations}\label{sec:derivations}
In this Appendix, we derive the steady-state equations for conservation of mass (eq.~[\ref{eq:Mdot}]), energy (eq.~[\ref{eq:Edot}]), and momentum (eq.~[\ref{eq:momentum}]) from the time-dependent equations of radiation hydrodynamics in a Schwarzschild metric \citep{Park2006}. We replace the notation of \cite{Park2006} with ours, for symbols that we have previously defined. We recover c.g.s. units by adding the necessary factors of $c$, and remove the angular terms to consider the spherically symmetric case. The equations are:
\begin{align}
    \frac{1}{\zeta^2}\frac{\partial}{\partial t}(n\Psi) + \frac{1}{r^2}\frac{\partial}{\partial r}(r^2nv\Psi) = 0 \qquad &\quad \text{Continuity equation}\label{eq:park1}\\
    \frac{\Psi}{\zeta^2}\frac{\partial}{\partial t}(v\Psi) + \frac{1}{2}\frac{\partial}{\partial r}(v\Psi)^2 + \frac{GM}{r^2} +  \frac{v\gamma^2}{\omega_g}\frac{\partial P_g}{\partial t} +
    \frac{c\Psi^2}{\omega_g}\frac{\partial P_g}{\partial r}
    = \frac{\Psi c}{\omega_g}\bar{\chi}\co F\co \qquad &\quad \text{Momentum equation}\label{eq:park2}\\
    \frac{n\Psi}{\zeta^2}\frac{\partial}{\partial t}\left(\frac{\omega_g}{n}\right) + nv\Psi\frac{\partial}{\partial r}\left(\frac{\omega_g}{n}\right) - \frac{\Psi}{\zeta^2}\frac{\partial P_g}{\partial t} - v\Psi\frac{\partial P_g}{\partial r} =  \Gamma\co-\Lambda\co \qquad &\quad \text{Energy equation}\label{eq:park3}\\
    \frac{1}{\zeta^2}\frac{\partial E\fx}{\partial t} + \frac{1}{\zeta^2r^2}\frac{\partial}{\partial r}(r^2\zeta^2F\fx) =  \frac{\gamma}{\zeta}(\Lambda\co-\Gamma\co-\frac{v}{c}\bar{\chi}\co F\co) \qquad &\quad \text{Zeroth radiation moment equation}\label{eq:park4}
\end{align}
In these equations, $n$ is the number density of particles, $\omega_g\equiv\rho c^2+P_g+U_g$ is the sum of rest-mass energy and enthalpy of the gas and $\bar{\chi}$ is the mean opacity coefficient and $E$ is the energy density of the radiation. $\Lambda$ and $\Gamma$ are heating and cooling functions that describe the interaction between the radiation and the gas. However, we will see that we can get rid of them to obtain our final equations. \citet{Park2006} also derives higher order moment equations of radiation, however we do not need them for this work since we consider a spherically symmetric problem, and we have a standalone equation for the radiation flux (eq.~[\ref{eq:FLD_flux}]).

In GR, the notion of frames of reference is important, and two were used by \citet{Park2006} to derive these equations; the fixed frame (subscript ``fx''), which has no velocity with respect to the origin $r=0$, and the comoving frame (subscript ``co''), which travels with velocity $u$, along with the gas. A frame transformation is used to convert between quantities in both frames (see \citet{Park2006} for the full details). What matters to us is that we are able to assign expressions to thermodynamic quantities in the comoving frame. The mean opacity coefficient in the frame of the moving gas can be related to the usual opacity, $\kappa\equiv \bar{\chi}\co/\rho$, since $\rho$ is measured in the frame. The flux that we keep track of is always the comoving flux, so we define $F\equiv F\co$. The local energy density is just $E\co\equiv U_R=aT^4$, for thermal radiation. The comoving radiation pressure in the radial direction is $P^{rr}\co\equiv P_R$, which is a function of the energy density $U_R$ and the flux limiter $\lambda$ (eq.~[\ref{eq:radiation_pressure}]).

We will now derive the steady-state equations. We will use the prime symbol ($'$) to denote derivatives with respect to $r$. From the continuity equation \eqref{eq:park1}, it is easy to see that $r^2nv\Psi$ is the conserved quantity in steady-state, and we can switch $n$ for $\rho$ since both densities are linked by the (constant) particle mass. Adding a factor of $4\pi$ for spherical geometry leads to equation for conservation of mass and $\Mdot$ (eq.~[\ref{eq:Mdot}]). The momentum equation \eqref{eq:park2} can be written compactly by including both the $v'$ and $GM/r^2$ terms into $\Psi'$, leading to our Eq.~\eqref{eq:momentum}. 

For the steady-state energy equation, we combine Eq.~\eqref{eq:park3} and \eqref{eq:park4} to remove the $\Lambda\co$ and $\Gamma\co$ functions, giving
\begin{equation}
    n v\Psi^2\left(\frac{w}{n}\right)'+\frac{1}{r^2}\left(r^2\zeta^2F\fx\right)'-v\Psi^2P_g'+\frac{v\Psi}{c}\rho\kappa F=0\,.
\end{equation}
Adding $v\Psi^2$ times Eq.~\eqref{eq:momentum} gets rid of the $P_g'$ and $F$ terms. We use mass conservation written as $(r^2nu\Psi)'=0$ to remove the $n'$ term, giving
\begin{equation}\label{eq:bernoulli}
    0=\frac{1}{r^2}\left(r^2\Psi^2vw+r^2\zeta^2F\fx\right)'\,.
\end{equation}
We have arrived at a Bernoulli equation for the flow, where the energy in the steady-state is a balance of radiation ($F\fx$), and rest mass, gravitational, kinetic and internal energies ($\Psi w$). To see this, we can expand $\Psi$ to first order,
\begin{equation}
    \Psi w \approx \left(1-\frac{GM}{c^2r}\right)\left(1+\frac{1}{2}\frac{v^2}{c^2}\right)(\rho c^2+P_g+U_g)
    \approx \rho\left(c^2-\frac{GM}{r}+\frac{v^2}{2}+\frac{P_g+U_g}{\rho}\right)\,,\label{eq:bernoulli_approx}
\end{equation}
where we ignored cross-products of small terms. Notice that the quantity in parentheses in Eq.~\eqref{eq:bernoulli_approx} is the usual non-relativistic Bernoulli's constant for an ideal gas in a gravitational potential. Now to obtain the integration constant, we integrate Eq.~\eqref{eq:bernoulli} and use the frame transformation for $F\fx$, giving
\begin{equation}
    C=r^2\Psi^2v\rho\left(\frac{w+U_R+P_R}{\rho}\right)+\Psi^2\left(1+\frac{v^2}{c^2}\right)r^2F\,.
\end{equation}
Then, in Eq.~\eqref{eq:Edot}, $\Edot\equiv4\pi C$ is the energy-loss rate.

\section{Parameter Spaces}\label{sec:parameter_spaces}
\begin{figure}
    \includegraphics{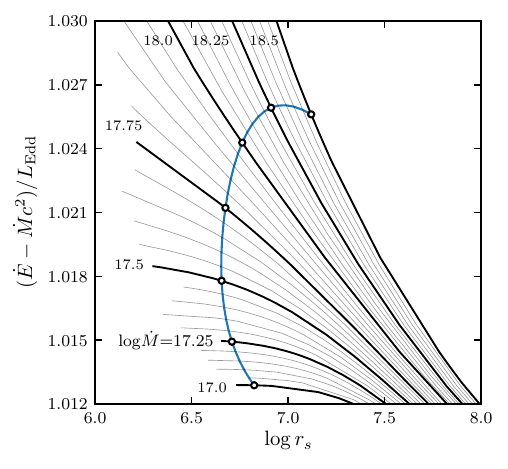}
    \includegraphics{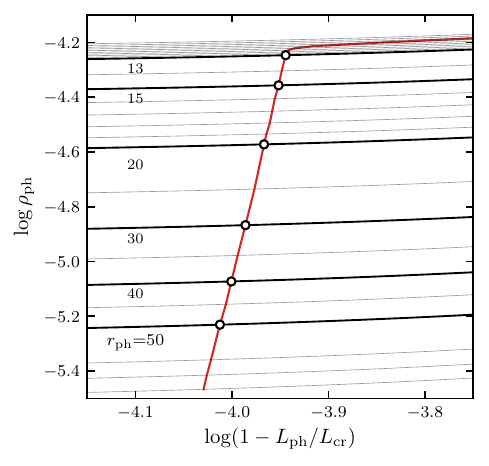}
    \caption{Parameter spaces for the wind (left) and envelope (right) models. The black lines trace points which allow a transition to optically thin and non-diverging integration to infinity. The thicker black lines are labelled by the mass-loss rates (in g s$^{-1}$) for winds, and the photospheric radii (in km) for envelopes. The blue and red lines trace out the points which satisfy the surface boundary condition.}
    \label{fig:param_spaces}
\end{figure}

We illustrate in Fig.~\ref{fig:param_spaces} the parameter spaces that we search to make wind and envelope models. In our code, we first perform a grid search to find which values allow integration to infinity, which yields the black lines shown in both panels. Then, we search along these lines to find which point satisfies the surface boundary condition when integrating inwards. Note that in practice, this numerical integration to infinity is not possible with a simple shooting method because of exponential growth in the stiff fluid equations \citep{Turolla1986}. To integrate our solutions outwards, we always kept track of two separate solutions with similar initial values (equal to 1 part in $10^4$), such that the two solutions would eventually diverge in opposite directions at some point in the integration. At that point, we interpolated values of $\rho$, $T$, $v$ between the two initial solutions to restart the integration with new initial values, pushing the divergence to larger radii. We kept doing the same process until we reached $r_\text{max}=10^9\; \rm{cm}$. At the end, we verified that our models satisfied the equations of structure (eq.~[\ref{eq:dvdr}]--[\ref{eq:dTdr}]) by plugging values back in and comparing derivatives. This method turned out to be very useful, and we were also able to use it for some of the inwards integrations to the neutron star surface that had similar stiffness issues.

In the left panel of Fig.~\ref{fig:param_spaces}, we represented the $\Edot$ free parameter as $\Edot-\Mdot c^2$ to look at the energy-loss rate without the rest mass contribution. What we see is that the remaining contributions to the energy-loss rates are roughly constant and just above $\Ledd$. To explain this, we can take the energy conservation equation \eqref{eq:Edot} and evaluate it at infinity,
\begin{align}
    &\Edot=\Linf\gamma_\infty^2\left(1+v_\infty^2/c^2\right)+\Mdot\gamma_\infty 
    \left(c^2+w_\infty\right)\,,\\
    &\Edot-\Mdot c^2\approx \Linf\gamma_\infty^2(1+v_\infty^2/c^2) +
    \Mdot c^2(\gamma_\infty-1)+ 2L^\infty\gamma_\infty v_\infty/c\label{eq:Edot-Mc2_approx}\,,
\end{align}
where the ``$\infty$'' subscript indicates evaluation at infinity for all variables. For the enthalpy at infinity $w_\infty$, we took the same estimate as in Eq.~\eqref{eq:enthalpy_infinity}. Eq.~\eqref{eq:Edot-Mc2_approx} shows that $v_\infty$ essentially dictates the small variance of $\Edot-\Mdot c^2$ in Fig.~\eqref{fig:param_spaces}. Indeed, in the first term, $\Linf{\sim} \Ledd$ for all models, and the last term can be neglected since, naturally, $T_\infty\rightarrow 0$. All that is left are terms of $v_\infty$. This is consistent with the velocity profiles in Fig.~\ref{fig:profiles}, where we see that the velocity asymptotically tends to higher values as $\Mdot$ increases,
except at the very high end of mass-loss rates, where it then begins to decrease.

For winds, numerical difficulties arise at the low end of the mass-loss rate, and we stopped our parameter space exploration at log$\Mdot=16.8$. This is because the sonic point radius quickly approaches the photosphere as $\Mdot$ decreases, which can be seen in the middle panel of Fig.~\ref{fig:triple} (this was also the case in \citet{Paczynski1986b}). This makes the outer integration extremely sensitive to the exact value of $r_s$ and $\Edot$, which can be seen in the left panel of Fig.~\ref{fig:param_spaces} with the lines of acceptable values becoming nearly flat.

In the right panel of Fig.~\ref{fig:param_spaces}, we plotted against $q\ph\equiv(1-L\ph/\Lcr)$ instead of $\Linf$ for better visualization. Note that there is a direct mapping between the two factors since $L\ph\equiv L(r\ph)=\Linf \zeta(r\ph)^{-2}$ and $\Lcr$ is a function of $r\ph$ and $T\ph$, which is itself a function of $\Linf$ with Eq.~\eqref{eq:env_Tph}. The values of $q\ph$ that we obtain for extended envelopes are much smaller than those of \citet{Paczynski1986a} (see their Fig.~2), likely because of our different treatment of the photosphere. For very compact envelopes, discussed in \S\ref{subsec:touchdown}, $q\ph$ becomes much larger, as can be seen in Fig.~\ref{fig:param_spaces}. Static atmospheres generally become convective as $L$ approaches $\Lcr$ \citep{Joss1973}. However, in our case the values of $\beta=P_g/P$ are sufficiently low that convection is avoided (see Fig.~39 of \citealt{Paxton2013}), as discussed by \cite{Paczynski1986a} for extended envelopes.

\section{Pressure conditions at the wind base}\label{sec:base_conditions}

In this Appendix we derive the ratio of gas pressure to total pressure $\beta$ expected in the initial stages of a burst. At constant pressure or column depth, nuclear burning generating an energy $E_\text{nuc}$ will raise the temperature of matter from $T_0$ to $T$ according to
\begin{equation}\label{eq:Enuc}
    E_\text{nuc} = \int_{T_0}^Tc_PdT\,,
\end{equation}
where $c_P$ is the heat capacity. For a mixture of ideal gas and radiation, $c_P=(5k_B/2\mu m_p)f(\beta)$ where $f(\beta)=(32-24\beta-3\beta^2)/5\beta^2$ \citep{Clayton1983}. Note that for $\beta\sim 1$ (gas pressure dominates), $f(\beta)\approx 1$ giving the standard ideal gas result for $c_P$. When $\beta\ll 1$, $f(\beta)$ diverges because fixing pressure also fixes temperature for a photon gas. The diverging heat capacity limits the final temperature to $T<T_R$, where $T_R$ is the radiation-pressure-limited temperature given by $gy = aT_R^4/3$ or or $T_R = 1.59\times 10^9\ {\rm K}\ (g_{14}/1.6)^{1/4}y_8^{1/4}$.

We make the change of variables $\eta\equiv1-\beta=aT^4/3P$, where $P=gy$ in hydrostatic equilibrium. $E_\text{nuc}$ is large enough that we can assume $T\gg T_0$ and Eq.~\eqref{eq:Enuc} can be re-written as
\begin{equation}\label{eq:Enuc2}
    E_\text{nuc}= \frac{5}{8}\frac{k_BT_R}{\mu m_p}\int_0^\eta \frac{5+30\eta-3\eta^2}{\eta^{3/4}(5-10\eta+5\eta^2)}d\eta = \frac{1}{2}\frac{k_BT_R}{\mu m_p}\frac{\eta^{1/4}(3\eta+5)}{1-\eta}.
\end{equation}
In the limit where $\beta=1-\eta \ll 1$, the expected value of $\beta$ is $\beta\approx 4k_BT_R/\mu m_p E_{\rm nuc}\approx 0.06g_{14}^{1/4}y_8^{1/4}\mu^{-1}E_{\text{nuc},18}^{-1}$, which is the expression used in \S\ref{subsec:masslossrate}.

\section{Analytical Newtonian envelopes}\label{sec:appendix_newtonian}

\citet{Paczynski1986a} showed a simple calculation for the most extended envelope in Newtonian gravity, one for which the luminosity ratio $\Gamma\equiv L/L_\text{Edd}=1$. Here we extend this calculation to the general case $\Gamma\leq 1$. With no general relativistic corrections, the hydrostatic balance and photon diffusion equations are simply written as
\begin{align}
	&\frac{dP}{dr}=-\frac{GM\rho}{r^2} \qquad ; \qquad
	\frac{dP_R}{dr}=-\frac{\rho\kappa L}{4\pi r^2c}\,.\label{eq:dPR_dr}
\end{align}
This leads to
\begin{equation}\label{eq:dPrdP}
	\frac{dP_R}{dP}=\frac{L}{L_\text{cr}}=\Gamma\frac{\kappa}{\kappa_0}=\Gamma\left[1+\left(\frac{T}{T_0}\right)^\alpha\right]^{-1}\,,
\end{equation}
where $T_0=4.5\times10^8$ K and $\alpha=0.86$ are from the opacity formula Eq.~\eqref{eq:kappa}. We may re-write Eq.~\eqref{eq:dPrdP} as 
\begin{equation}
	dP=\frac{1}{\Gamma}\frac{4a}{3}\left[1+\left(\frac{T}{T_0}\right)^\alpha\right]T^3dT
\end{equation}
which we integrate from the photosphere $r_\text{ph}$ where we assume $T\approx 0$ and thus $P\approx 0$, giving the general expression
\begin{equation}
	P(T) = \frac{1}{\Gamma}\frac{aT^4}{3}\left[1+\frac{4}{4+\alpha}\left(\frac{T}{T_0}\right)^\alpha\right]\,.
\end{equation}
This also leads to an expression for the density, since $P_g=P-P_R=kT\rho/\mu m_p$, such that
\begin{equation}
	\rho(T)=\frac{1}{\Gamma}\frac{\mu m_p}{k}\frac{aT^3}{3}\left[1-\Gamma+\frac{4}{4+\alpha}\left(\frac{T}{T_0}\right)^\alpha\right]\label{eq:newt_rho}
\end{equation}
Putting this back into Eq.~\eqref{eq:dPR_dr}, we obtain a differential equation for $T$,
\begin{equation}
	\left[1+\left(\frac{T}{T_0}\right)^\alpha\right]\left[1-\Gamma+\frac{4}{4+\alpha}\left(\frac{T}{T_0}\right)^\alpha\right]^{-1}dT=-\frac{1}{4}\frac{\mu m_p}{k}\frac{GM}{r^2}dr \,.
\end{equation}
This can be integrated from the photosphere. The $\Gamma=1$ case is straightforward and leads to the expression in \citet{Paczynski1986a},
\begin{equation}
	\frac{GM}{r}\frac{\mu m_p}{kT}\frac{1}{4+\alpha}\left(1-\frac{r}{r_\text{ph}}\right)=1+\frac{1}{1-\alpha}\left(\frac{T_0}{T}\right)^\alpha \,.
\end{equation}
If $\Gamma<1$, we instead have 
\begin{align}
	&\frac{GM}{r}\frac{\mu m_p}{kT}\frac{1}{4+\alpha}\left(1-\frac{r}{r_\text{ph}}\right)\nonumber\\
	&=1-\left(1-\frac{4}{(4+\alpha)(1-\Gamma)}\right){}_2F_1\left(1,\frac{1}{\alpha};1+\frac{1}{\alpha};\frac{-4(T/T_0)^\alpha}{(4+\alpha)(1-\Gamma)}\right)\,,
\end{align}
where ${}_2F_1$ is the hypergeometric function.  All that is required to find $r_\text{ph}$ for a given $\Gamma$ is to have a known pair $(r,T)$ somewhere in the envelope. For example, \citet{Paczynski1986a} assumed a constant $T=2\times10^9$ K at $r=R$. For consistency, we use our boundary condition $P=gy_b$ with $y_b=10^8\ {\rm g\ cm^{-2}}$ at $r=R$, which we can easily solve for $T$ since Eq.~\eqref{eq:newt_rho} gives $\rho=\rho(T)$. This is how we computed the Newtonian envelope models shown in Fig.~\ref{fig:touchdown}.

\end{document}